\documentclass[12pt]{iopart}

\usepackage{bm,graphicx,iopams} 
\newcommand{\nn}{\nonumber}

\begin{document}

\title[Approximate Method of Variational Bayesian MF/MC with Sparse Prior]{Approximate Method of Variational Bayesian Matrix Factorization/Completion
 with Sparse Prior}

\author{Ryota Kawasumi and Koujin Takeda}

\address{Department of Intelligent Systems Engineering, Graduate School of Science and Engineering, Ibaraki University}
\ead{koujin.takeda.kt@vc.ibaraki.ac.jp}
\vspace{10pt}

\begin{abstract}
 We derive analytical expression 
 of matrix factorization/completion solution by variational Bayes method, under the assumption
 that observed matrix is originally the product of low-rank
 dense and sparse matrices with additive noise. 
 We assume the prior of sparse matrix is Laplace distribution
 by taking matrix sparsity into consideration.
 Then we use several approximations for derivation of matrix factorization/completion solution.
 By our solution, we also numerically evaluate the performance of sparse matrix reconstruction in matrix 
 factorization, and completion of missing matrix element in matrix completion.
\end{abstract}

%
%
%
%
%


\section{Introduction}
{\it Matrix factorization} (MF) is the problem of factorizing single matrix
into two low-rank matrices. 
{\it Matrix completion} (MC) is also a problem for matrix, 
where we must infer unobserved matrix elements
using low-rank assumption like MF.
By solving MF/MC problem, we can deal with various data analysis problems like principal component analysis or 
collaborative filtering. 

In this article we study the problem of MF/MC with additive noise. 
Many algorithms are proposed and used for MF/MC \cite{SRJ, CP, CCS}, and here
we make Bayesian inference approach based on statistical physics. 
In solving the problem by Bayesian formulation, 
posterior probability is a complicated function and intractable in general.
One of the standard approaches to cope with this difficulty is 
Markov chain Monte Carlo method, which has already been used for MF in \cite{SM}. 
Another approach is message passing or equivalently cavity formulation,
which is found in \cite{KKMSZ} and applicable to many problem settings in MF
as mentioned there. 

We use variational Bayes (VB) method to manage
above-mentioned difficulty analytically.
We consider Kullback-Leibler (KL) divergence between true posterior 
and trial function, and attempt to find the trial function to minimize it. 
In the preceding work, VB has already been used for MF \cite{NS,NSBT}, where
they assumed multivariate Gaussian prior with zero mean vector and 
diagonal covariance matrix. Under these assumptions, they obtained analytical solution 
of global KL divergence minimum with the aid of singular value decomposition.

In the present paper we assume one of low-rank matrices is sparse,
whose assumption is supposed to deal with dictionary learning \cite{OF1,OF2}. 
For dictionary learning, several algorithms have also been proposed \cite{EAH,AEB,ZT},
and the performance of dictionary determination 
is discussed by statistical mechanical method \cite{SK,KMZ,KKMSZ}.

For sparse factorized matrix, we use Laplace prior for convenience of VB analysis.
Then our goal is to find the analytical expression of moments of
trial function in VB, which minimizes KL divergence between true posterior and trial function itself.
However, we must use several approximations due to the reasons as follows.
First, we cannot perform multiple integrals in
the evaluation of the moments of trial functions 
without approximations. Second, we need to simplify the result for 
avoiding intractability in numerical experiment. 
With them, we obtain the analytical expression of KL divergence minimum
as the relations between the first and the second moments of trial functions
for two factorized matrices.
Then we regard our analytical result as an iterative algorithm of MF/MC, and 
evaluate the performance of MF/MC by numerical experiment. 
The result validates the assumption of Laplace prior for the current problem.

This paper is organized as follows. We give the definition of the model and 
Bayesian formulation for MF/MC in section 2. We briefly review VB method
and elucidate how to apply VB to MF/MC in section 3. Several approximations for VB 
are introduced in section 4. Our analytical result is summarized in section 5. 
We show the result of numerical experiment by the algorithm based on our analytical result 
in section 6. 
Section 7 is devoted for summary and future perspective.

\section{Model}
Throughout this article we describe matrix or vector by boldface letter.
The element in matrix or vector is denoted by the regular letter with subscript 
(e.g. $x_{lh}$ for $lh$-element in $\bm X$).
Each subscript represents the position in matrix or vector, and takes the values depending on the characters
 as follows: 
$l, l', l'' \in \{1, 2, \cdots, L\}$, $m, m', m'' \in \{1, 2, \cdots, M\}$, $h, h', h'' \in \{1, 2, \cdots, H\}$.

In MF, the observed matrix ${\bm V} \in {\mathbb R}^{LM}$ 
has no missing element, and is described by the product of two matrices
${\bm A} \in {\mathbb R}^{LH} $ and
${\bm B} \in {\mathbb R}^{HM} $ with additional noise matrix
${\bm E} \in {\mathbb R}^{LM}$,
\begin{eqnarray}
{\bm V}
&=&
{\bm A}{\bm B} + {\bm E}.
\end{eqnarray}
In MC, some of observed matrix elements $\bm V$ are missing, which is expressed as
\begin{eqnarray}
{\bm V}
&=&
\bm \Theta \circ (
{\bm A}{\bm B} + {\bm E}).
\end{eqnarray}
The symbol $\circ$ means Hadamard or element-wise product.
The element in $\bm \Theta \in \mathbb{R}^{LM}$ is 
unity for observed element and zero for unobserved.
Therefore, both of fully-observed and partially-observed cases are included
in the formulation.

Our task is to determine matrices ${\bm A}, {\bm B}$ from ${\bm V}$.
We assume that the prior of ${\bm A}$ is multivariate Gaussian 
with zero mean vector and covariance matrix ${\bm C}_{\bm A} \in \mathbb{R}^{H^2}$.
The covariance matrix has only diagonal element for simplicity and
later convenience of numerical experiment.
In contrast, we assume that the matrix ${\bm B}$ is sparser than ${\bm A}$, 
and we use Laplace distribution with zero mean and 
variance $2k^2$ for sparse prior.
To summarize,
\begin{eqnarray}
P({\bm A}) 
&\propto&
\exp \left( - \frac{1}{2} {\rm tr} \bm A^T (\bm C_{\bm A})^{-1} \bm A \right) = 
\prod_{l,h} \exp\Bigr( - \frac{ a_{lh}^2  }{2({C}_{\bm A})_{hh} } \Bigr),
\label{P1}
\\
P({\bm B})
 &\propto&
\prod_{h,m}
\exp\Bigr( - \frac{ |b_{hm}|  }{ k } \Bigr).
\label{P2}
\end{eqnarray}
The probability distribution of the element in $\bm E$ is
i.i.d. Gaussian with
zero mean and variance $\sigma^2$.
Consequently, the likelihood can be expressed as
\begin{eqnarray}
P({\bm V}|{\bm A}, {\bm B})
&\propto&
\prod_{l,m}
\exp\Bigr\{ - \frac{1}{2\sigma^2} \theta_{lm} \Bigr( v_{lm} - \sum_h a_{lh}b_{hm} \Bigr)^2   \Bigr\}.
\label{P3}
\end{eqnarray}

\section{Variational Bayes Method}

Our aim is to find the solution of factorized matrices $\bm A, \bm B$ by Bayesian inference. 
The standard approach is to maximize posterior $P(\bm A, \bm B| \bm V)$ with respect
to $\bm A, \bm B$ for a given observed matrix $\bm V$. (We omit the information of the missing matrix
$\bm \Theta$ in the Bayesian formulation.) 
The posterior ${\it P}({\bm A}, {\bm B}|{\bm V})$ is expressed by
the priors ${\it P}({\bm A}), {\it P}({\bm B})$ and the likelihood
${\it P}({\bm V}|{\bm A}, {\bm B})$ using Bayes formula,
\begin{eqnarray}
P({\bm A},{\bm B}|{\bm V})
&=&\frac{P({\bm V}|{\bm A},{\bm B})P({\bm A})P({\bm B})}
{\int_{\mathbb{R}^{LH+HM}} P({\bm V}|{\bm A},{\bm B})P({\bm A})P({\bm B}) d{\bm A}d{\bm B}},
\label{Bayesformula}
\end{eqnarray}
where we simplify the symbol of multiple integrals as single.
However, the solution of posterior maximization is generally difficult to find.
This difficulty is equivalent to the fact that the estimation of the denominator, 
or partition function in statistical physics, is hard to calculate
both analytically and computationally.

To avoid such difficulty, we attempt to estimate 
the factorized matrices ${\bm A}, {\bm B}$ analytically by 
VB method, or 
KL divergence minimization by variational calculus \cite{ref9}.
KL divergence between two probability distributions $p (x), q (x)$ is defined by
\begin{eqnarray}
{\rm KL}(p \| q)&:=&-{\int p({x})\ln{\frac{q({x})}{p({x})}}d{x}}.
\end{eqnarray}
We prepare the trial function 
$r({\bm A},{\bm B})$ and determine 
it by the condition that KL divergence between true posterior and trial function is minimum.
Then, we express KL divergence as a functional with respect to $r({\bm A}, {\bm B})$,
\begin{eqnarray}
F(r)=
{-\int_{\mathbb{R}^{LH}} d{\bm A} \int_{\mathbb{R}^{HM}} d{\bm B} \ r({\bm A},{\bm B}) \ln{\frac{P({\bm A},{\bm B}|{\bm V})}{r({\bm A},{\bm B})}}} +\ln{p({\bm V})},
\end{eqnarray}
where $p(\bm V)$ is the denominator in (\ref{Bayesformula}).
The functional $F(r)$ corresponds to variational free energy in physics.
For convenience, we assume that the trial function is statistically independent with respect to 
the matrices $\bm A$ and $\bm B$,
\begin{equation}
r({\bm A}, {\bm B}) = r({\bm A}) r({\bm B}).
\end{equation}
With this, the trial functions $r({\bm A}), r({\bm B})$ minimizing KL
can be obtained. They allow us to infer the factorized matrices ${\bm A}, {\bm B}$
by their mean values,
\begin{eqnarray}
\bar{a}_{lh} := \int_{\mathbb{R}^{LH}} a_{lh} r(\bm A) d \bm A, \label{meana}\\
\bar{b}_{hm} := \int_{\mathbb{R}^{HM}} b_{hm} r(\bm B) d \bm B. \label{meanb}
\end{eqnarray}
As stated in \cite{NW},
for arbitrary priors $P(\bm A), P(\bm B)$ and likelihood $P(\bm V|\bm A,\bm B)$,
trial functions $r({\bm A}), r({\bm B})$ minimizing KL 
are exactly evaluated by variational calculus with the constraint of normalization condition for trial functions.
The result is
\begin{eqnarray}
r({\bm A})
&\propto&
P({\bm A})\exp\Bigr\{ \int_{\mathbb{R}^{HM}} r({\bm B}) \ln P({\bm V}|{\bm A}, {\bm B}) d{\bm B}  \Bigr\},
\label{r1}
\\
r({\bm B})
&\propto&
P({\bm B})\exp\Bigr\{ \int_{\mathbb{R}^{LH}} r({\bm A}) \ln P({\bm V}|{\bm A}, {\bm B}) d{\bm A}  \Bigr\}.
\label{r2}
\end{eqnarray}

\section{Approximation}

However, we still have difficulty in calculation of means of trial functions $r(\bm A), r(\bm B)$
in (\ref{meana}) and (\ref{meanb}), because we need to perform high dimensional integral
with respect to all elements in $\bm A$ or $\bm B$.
We avoid it by the approximations as below.
 
\begin{enumerate}
\renewcommand{\labelenumi}{(\roman{enumi})}
\item{mean field approximation:\\

After substitution of equations (\ref{P1})-(\ref{P3})
into (\ref{r1}) and (\ref{r2}), we find 
it difficult to perform multiple integrals. 
Therefore, we use mean field approximation,
namely all elements in ${\bm A}, {\bm B}$ are mutually statistically independent.
With it, we can integrate out variables, which are not concerned in 
the evaluation of means in (\ref{meana}) and (\ref{meanb}),
\begin{eqnarray}
r({\bm A})r({\bm B})
&=&
\prod_{l,h} r_a(a_{lh}) \prod_{h,m}r_b(b_{hm}).
\end{eqnarray}
}

\item{ large $k$ approximation up to the first order of $1/k$:\\

Even with mean field approximation, we still have problem in
calculation of integral with Laplace prior. Due to Laplace prior,
we need to separate the integration range into positive and negative 
for each matrix element in $\bm B$,
which yields the summation over $2^{HM}$ terms
after integration.
In principle, we can write down the exact result of integration. 
However the final expression has exponential number of terms as mentioned, 
which leads to the difficulty in the inference of factorized matrices $\bm A, \bm B$.
For this reason, we take the limit of $k \to \infty$, and keep only 
the terms up to $O(1/k)$ in Laplace prior,
\begin{eqnarray}
P({\bm B})
&=&
\prod_{h,m}
\exp\Bigr( - \frac{ |b_{hm}|  }{ k } \Bigr)
\ \ \stackrel{k \to \infty}{\longrightarrow} \ \ 
1 - \sum_{h,m}\frac{|b_{h m}|}{k}.
\end{eqnarray}
Here $b_{hm} \in \mathbb{R}$ and not limited within the range of $P(\bm B) \ge 0$. 
Then we must keep in mind that probability may become negative in smaller $k$ region by this approximation,
which may arise another problem in the evaluation of $\bm A, \bm B$.
The validity of this approximation should be
checked later.\\
}

\item{diagonal covariance matrix of $r({\bm A}), r({\bm B})$:\\

In the past work using Gaussian prior \cite{NS,NSBT}, 
the full covariance matrix is taken into consideration, 
while we use only variance and off-diagonal element 
in covariance matrix (i.e. covariance component) is set to be zero.

As stated in \cite{NS}, in the case of Gaussian priors for $\bm A, \bm B$, 
we can determine the form of trial functions $r(\bm A), r(\bm B)$ minimizing KL divergence, 
which are multivariate Gaussians. 
Then we can write down the expression of minimal KL divergence explicitly.
Using this, it is shown that off-diagonal elements can be ignored 
for the discussion of KL divergence minimization without loss of generality. 
On the other hand, in Laplace prior case we cannot determine the form of trial functions. 
Hence, for the present we use diagonality of covariance matrix 
as an {\it approximation} to follow the analysis of Gaussian case. 
If nonzero off-diagonal covariance matrix elements are introduced, 
the analysis will be much complicated, and we leave it as a future problem.
Furthermore, in our formulation
we use only the first and the second moments of $r(\bm A), r(\bm B)$
and neglect higher, 
which is also an approximation.

Under these approximations, the mean and the variance of ${\bm A}, {\bm B}$ are calculated as
\begin{eqnarray}
\overline{a}_{lh} &=& \int_{\mathbb{R}} a_{lh} r_a(a_{lh}) da_{lh},
\\
\overline{b}_{hm} &=& \int_{\mathbb{R}} b_{hm} r_b(b_{hm}) db_{hm},
\\
({\Sigma}_{{\bm A} l})_{hh}
&=&
\int_{\mathbb{R}} a_{lh}^2 r_a(a_{lh}) d a_{lh} -\overline{a}_{lh}^2,
\\
({\Sigma}_{{\bm B} m})_{hh}
&=&
\int_{\mathbb{R}} b_{hm}^2 r_b(b_{hm}) d b_{hm} -\overline{b}_{hm}^2.
 \end{eqnarray}
}
\end{enumerate}
Using these approximations, our final objective is to obtain relations between the first and the second
moments of the trial functions.

\section{Result of VB analysis}

\subsection{Final expression of KL divergence minimum}
As the result of KL divergence minimization by VB, we obtain
the expression of equations for means and variances of $\bm A, \bm B$
as shown below. See appendix for the detail of the derivation.
\begin{eqnarray}
\bar{a}_{lh}
&=&
\sum_{m,h'}\theta_{lm} 
(\hat{\Sigma}_{{\bm A}l}^{-1})_{hh'} v_{lm} \bar{b}_{h'm},
\label{A1} \\
({\Sigma}_{{\bm A} l} )_{hh}
&=&
\sigma^2 (\hat{\Sigma}_{{\bm A}l}^{-1})_{hh},
\label{A2} 
\end{eqnarray}
\begin{eqnarray}
\bar{b}_{hm}
&=&
\sum_{h',l} \theta_{lm} (\hat{\Sigma}_{{\bm B} m}^{-1})_{hh'} v_{lm} \bar{a}_{lh'} 
-
\sum_{h'}
\frac{ \sigma^2(\hat{\Sigma}_{{\bm B}m}^{-1})_{h' h}  }{k Z_B}
{\rm erf}\bigr(\omega_{h' m}\bigr),\label{B1} 
\end{eqnarray}
\begin{eqnarray}
\nn \\
(\Sigma_{{\bm B}m})_{hh}
&=&
\sigma^2 (\hat{\Sigma}_{{\bm B}m}^{-1})_{hh} \nn \\
&&-
\sum_{h' }
\sqrt{\frac{2}{\pi \sigma^2 (\hat{\Sigma}_{{\bm B}m}^{-1})_{h'h'}}}
\frac{ \{\sigma^2 (\hat{\Sigma}_{{\bm B}m}^{-1})_{h' h} \}^2  }{ k Z_B  }
\exp\bigr(-\omega_{h' m}^2\bigr)
\nn \\
&& -
\Bigr\{
\sum_{h'}
\frac{ \sigma^2(\hat{\Sigma}_{{\bm B}m}^{-1})_{h' h} }{k Z_B }
{\rm erf}\bigr(\omega_{h' m}\bigr)
\Bigr\}^2,
\label{B2}
\end{eqnarray}
where the definition of error function is ${\rm erf}(x) := (2/\sqrt{\pi}) \int_x^{\infty} \exp(-t^2) dt$,
 and
\begin{eqnarray}
\omega_{hm} 
&=&
\frac{\sum_{h',l} \theta_{lm} (\hat{\Sigma}_{{\bm B}m}^{-1})_{hh'} 
v_{lm} \bar{a}_{lh'} }{\sqrt{2 \sigma^2 (\hat{\Sigma}_{{\bm B}m}^{-1})_{hh} }},  
\label{def4}
\\
Z_B &=&
1 - \frac{1}{k} \sum_{h, m}
\left\{
\sqrt{ \frac{ 2 \sigma^2 (\hat{\Sigma}_{{\bm B} m}^{-1})_{hh} }{ \pi}  } 
\exp\bigr(-\omega_{h m}^2\bigr) \right. \nn \\
&& \hspace{2cm} \left. +
\Bigr(
\sum_{h',l} \theta_{lm} (\hat{\Sigma}_{{\bm B} m}^{-1})_{h h'} v_{lm} \bar{a}_{lh'}
\Bigr)
 {\rm erf} \bigr(\omega_{h m}\bigr) \right\}.
 \end{eqnarray}
$Z_B$ means the renormalization factor of the probability in the calculation of 
mean and variance of the matrix $\bm B$.
(See also appendix.)
The matrices $\hat{\bm \Sigma}_{{\bm A} l}, \hat{\bm \Sigma}_{{\bm B} m}
 \in {\mathbb R}^{H^2}$ are obtained as
\begin{eqnarray}
(\hat{ \Sigma}_{{\bm A}l})_{hh'}
&=&
\frac{\displaystyle \sigma^2}{\displaystyle ({ C}_{{\bm A}})_{hh}} \delta_{h,h'}
+\sum_m\theta_{lm}\Bigl(({ \Sigma}_{{\bm B} m})_{hh} \delta_{h,h'} +
\bar{b}_{hm}\bar{b}_{h'm}
\Bigr), \label{def6}\\
(\hat{\Sigma}_{{\bm B} m})_{hh'}
&=&
\sum_l\theta_{lm}\Bigl(({ \Sigma}_{{\bm A }l})_{hh}\delta_{h,h'}+\bar{a}_{lh}\bar{a}_{lh'}\Bigr),
\label{def5}
\end{eqnarray}
where $\delta_{h,h'}$ is Kronecker delta.

By solving these four equations (\ref{A1})-(\ref{B2}) in conjunction
with the evaluation of variables on l.h.s. of (\ref{def4})-(\ref{def5}),
we can infer the factorized matrices from the estimation
of $\bar{a}_{lh}, \bar{b}_{hm}$. 
It seems difficult to have simpler analytical expression 
of (\ref{A1})-(\ref{B2}) further,
therefore we resort to the numerical experiment using these expressions
as explained in section 6.

\subsection{Relation with the past work}
The limit of $k \to \infty$ with keeping only $O(1)$ term
is the case of uniform prior $P({\bm B}) = {\rm const.}$
In this case, the result of VB analysis 
is included in the past work of Gaussian prior \cite{NS,NSBT} by 
taking the limit of infinite variance. 
We verified that 
our analytical expression reproduces the uniform prior case for $k \to \infty$.
Accordingly, we can identify the correction term from Laplace prior.

The limit of $k \to \infty$ is expressed as
\begin{eqnarray}
\overline{a}_{lh}^{(k\rightarrow\infty)}
&=&
\sum_{m,h'}
(\hat{\Sigma}_{{\bm A} l}^{-1})_{hh'} v_{lm} \overline{b}_{h'm},
\label{A1kinfty}
\\
({\Sigma}_{{\bm A} l} )_{hh}^{(k\rightarrow\infty)}
&=&
\sigma^2 (\hat{\Sigma}_{{\bm A} l}^{-1})_{hh},
\label{A2kinfty}
\\
\overline{b}_{hm}^{(k\rightarrow\infty)}
&=&
\sum_{l,h'} (\hat{\Sigma}_{{\bm B} m}^{-1})_{hh'}
 v_{lm} \overline{a}_{lh'},
\label{B1kinfty}
\\
({\Sigma}_{{\bm B} m} )_{hh}^{(k\rightarrow\infty)}
&=&
\sigma^2 (\hat{\Sigma}_{{\bm B} m})_{hh}^{-1},
\label{B2kinfty}
\end{eqnarray}
for fully-observed case (namely $\theta_{lm}=1\  \forall l,m$), 
where the matrices 
$\hat{\bf \Sigma}_{{\bm A} l}, \hat{\bf \Sigma}_{{\bm B} m} \in {\mathbb R}^{H^2}$ are
given by (\ref{def6}) and (\ref{def5}).

These equations can readily be solved numerically, or analytically by the discussion given in \cite{NS}.
The discussion there is summarized as follows.
As mentioned, in Gaussian prior case 
it is shown that the minimum solution of KL divergence 
has diagonal covariance matrix, therefore we can consider diagonal ${\bf \Sigma}_{{\bm A} l}^{(k \to \infty)}$ and 
${\bf \Sigma}_{{\bm B} m}^{(k \to \infty)}$. 
Next, applying singular value decomposition to the 
matrix product $\bar{\bm A}^{(k \to \infty)} \bar{\bm B}^{(k \to \infty)}$,
we can express the matrices $\bar{\bm A}^{(k \to \infty)}$ and $\bar{\bm B}^{(k \to \infty)}$ 
by the left/right eigenvectors of the product $\bar{\bm A}^{(k \to \infty)} \bar{\bm B}^{(k \to \infty)}$. 
Using these results, the equations for $\bar{\bm A}^{(k \to \infty)}, \bar{\bm B}^{(k \to \infty)}, {\bf \Sigma}_{{\bm A} l}^{(k \to \infty)}, {\bf \Sigma}_{{\bm B} m}^{(k \to \infty)}$ are finally reduced to algebraic equations,
which can be solved analytically. 
As a consequence, the variables in (\ref{A1kinfty})-(\ref{B2kinfty}) are represented by the singular values and the left/right eigenvectors.

Furthermore, by this argument we can guarantee 
that the result in (\ref{A1kinfty})-(\ref{B2kinfty}) describes global minimum of KL divergence. 
On the other hand, for Laplace prior we use several approximations, and our final equations are nonlinear.
Therefore guarantee of global minimum is not clear.

\section{Numerical Experiment}
\subsection{Methodology of numerical experiment}

By solving equations (\ref{A1})-(\ref{B2}) numerically,
we can evaluate MF/MC performance of our analytical result.
Then, based on our result, we conduct numerical experiment using synthetic data.
For numerical experiment, we regard the equations (\ref{A1})-(\ref{B2}) as an iterative algorithm,
where $\bar{a}_{lh}$ and $\bar{b}_{hm}$ are determined alternately. 
After convergence of $\bar{a}_{lh}$ and $\bar{b}_{hm}$, 
the original matrices $\bm A, \bm B$ can be inferred from their values.
 
The method of numerical experiment is summarized as follows:

\begin{enumerate}
\renewcommand{\labelenumi}{(\roman{enumi})}
\item{{\it Initialization}:
\begin{enumerate}
\item{Prepare the original matrices 
${\bm A}\in {\mathbb R}^{LH}$ and ${\bm B'}\in {\mathbb R}^{HM}$,
whose element is randomly drawn from ${\cal N}(0,1)$. Prepare
noise matrix ${\bm E} \in {\mathbb R}^{LM}$ as well, 
whose element is randomly drawn from ${\cal N}(0, \sigma^2)$.}
\item{  Replace some elements in ${\bm B'}$ with zero for introducing sparsity.
The locations of zero elements are chosen randomly. 
The matrix after replacement is denoted by ${\bm B}$. 
}
\item{ For MC experiment, prepare the missing matrix $\bm \Theta$. 
The element is unity 
if the corresponding element in $\bm V$ is observed, otherwise zero.
The locations of missing elements are chosen randomly.
}
\item{ Compute the observed matrix ${\bm V}\in {\mathbb R}^{LM}$ by
${\bm V} = \bm \Theta \circ ({\bm A}{\bm B} + {\bm E})$.}
\item{ Initialize all elements $\bm A, \bm B$ by drawing
from ${\cal N}(0,1)$ randomly. The initial matrices are denoted by ${\bm A}^{(0)}, {\bm B}^{(0)}$, respectively.
(The superscript represents the iteration step.)
Initialize covariance matrices ${\bm \Sigma}_{\bm A l}^{(0)}, {\bm \Sigma}_{\bm B m}^{(0)}$
to be identity. Set the counter of iteration step $t=1$.\\}
\end{enumerate}
}

\item{{\it Iteration}:

In $t$-th step,
substitute the elements in ${\bm A}^{(t-1)}, {\bm B}^{(t-1)},$
${\bm \Sigma}_{\bm A l}^{(t-1)}, {\bm \Sigma}_{\bm B m}^{(t-1)}$ 
into r.h.s. of equations (\ref{A1})-(\ref{B2}).
(The means $\bar{a}_{lh}$ and $\bar{b}_{hm}$ on r.h.s. are replaced by
${a}_{lh}^{(t-1)}$ and $b_{hm}^{(t-1)}$, respectively.)
Then compute $\bar{a}_{lh}, 
\bar{b}_{hm}$ on l.h.s. of (\ref{A1}) and (\ref{B1}),
whose results are regarded as ${\bm A}^{(t)}$ and ${\bm B}^{(t)}$, respectively.
Similarly, compute ${\bm \Sigma}_{\bm A l}^{(t)}, {\bm \Sigma}_{\bm B m}^{(t)}$}
on l.h.s. of (\ref{A2}) and (\ref{B2}).\\

\item{{\it Convergence check}:

Check whether the convergence condition below holds or not
\begin{eqnarray}
&& \hspace{-1cm}
\Bigr\{
 d({\bm A}^{(t)} {\bm B}^{(t)}, {\bm A}{\bm B}) 
-d({\bm A}^{(t-1)} {\bm B}^{(t-1)}, {\bm A}{\bm B})
\Bigr\}^2 < \delta,
\end{eqnarray}
for a given small threshold value $\delta$, where 
\begin{equation}
d(\bm X, \bm Y) 
:= \|\bm X - \bm Y\|_{\rm FRO}^2 / \| \bm Y \|_{\rm FRO}^2
\end{equation}
is the squared normalized distance between matrices $\bm X, \bm Y$.
The symbol $\| \cdot \|_{\rm FRO}$ is Frobenius norm defined by $\| \bm X \|_{\rm FRO}
:= \sqrt{\sum_{lm} (x_{lm})^2}$.

Once the condition is satisfied, terminate the iteration,
record the final iteration step $n$,
and take ${\bm A}^{(n)}, {\bm B}^{(n)}$ as a result of MF/MC.
(For MC, reconstruct the matrix product by $\bm A^{(n)} \bm B^{(n)}$.)
Then go to (iv).
Otherwise, increase the counter of iteration step $t \rightarrow t+1$, and return to (ii).}\\

\item{{\it Performance evaluation}:\\
After convergence, we observe the difference between the original sparse matrix $\bm B$ and the reconstructed one
${\bm B}^{(n)}$ by computing the errors in (\ref{errB})-(\ref{errVMC}).
Because the solution of MF has degeneracy under simultaneous permutation of 
columns in $\bm A$ and rows in $\bm B$, the best $\bm B^{(n)}$ minimizing 
errors under permutation is chosen in the evaluation of errors, which are denoted by 
$\widetilde{\bm B}^{(n)}$ in (\ref{errB})-(\ref{errBsp}). 
We will comment on the degeneracy of solution again in subsection \ref{result3}.

First, we evaluate ${\rm err}{\bm B}$ in (\ref{errB}), i.e. the normalized mean squared error of $\bm B^{(n)}$
to the original $\bm B$ for $\it all$ matrix elements. 
We also evaluate ${\rm err}{\bm B}^{\rm abs}$ in (\ref{errBabs}), 
namely the normalized mean squared error after taking absolute values for all matrix elements.
This quantity is more suitable for the discussion of original matrix reconstruction, because
MF solution has another degeneracy under
sign inversion of the column in $\bm A$ and the paired row in $\bm B$, as again mentioned later.
(The symbol $|\bm B|$ in (\ref{errBabs}) means 
the matrix after taking absolute value of all matrix elements in $\bm B$.)
Next, for performance of sparse element identification,
evaluate ${\rm err}{\bm B}^{\rm sp}$ in (\ref{errBsp}), 
the normalized mean squared error only for {\it sparse} elements in the original matrix satisfying $b_{hm} = 0$.
Finally, for correctness of MF solution, evaluate ${\rm err} \bm A \bm B$ in (\ref{errV}), namely
the normalized error of the whole elements in the matrix product $\bm A \bm B$.
In addition, only for MC problem, evaluate the error for the missing elements in $\bm A \bm B$, 
denoted by ${\rm err} \bm A \bm B^{\rm MC}$ in (\ref{errVMC}).
\begin{eqnarray}
{\rm err}{\bm B}
&:=&
d \left(\frac{ \widetilde{\bm B}^{(n)}} { \|\widetilde{\bm B}^{(n)}\|_{\rm FRO} } ,
\frac{{\bm B}}{ \|{\bm B}\|_{\rm FRO} } \right),\label{errB} \\
{\rm err}{\bm B}^{\rm abs}
&:=&
d \left(\frac{ |\widetilde{\bm B}^{(n)}|}{ \|\widetilde{\bm B}^{(n)}\|_{\rm FRO} } ,
\frac{|{\bm B}|}{ \|{\bm B}\|_{\rm FRO} } \right),\label{errBabs} \\
{\rm err}{\bm B}^{\rm sp}
&:=&
{\displaystyle 
\sum_{ \tiny (h,m)\in \{ b_{hm}=0 \}  }}
\hspace{-1mm} \left\{ 
\frac{ \widetilde{b}_{hm}^{(n)}  }{ \|\widetilde{\bm B}^{(n)}\|_{\rm FRO} } 
\right\}^2, \label{errBsp} \\
{\rm err}{\bm A \bm B}
&:=&
d \left( \frac{ {\bm A^{(n)} \bm B}^{(n)}} { \| {\bm A^{(n)} \bm B}^{(n)}\|_{\rm FRO} },
\frac{{\bm A \bm B}}{ \|{\bm A \bm B}\|_{\rm FRO} } \right),\label{errV} \\
{\rm err}{\bm A \bm B}^{\rm MC}
&:=&
\sum_{ \tiny (l,m)\in \{ \theta_{lm}=0 \}  }
\left\{ \frac{ \sum_{h} {a}_{lh}^{(n)} {b}_{hm}^{(n)} } { \| {\bm A^{(n)} \bm B}^{(n)}\|_{\rm FRO} } -
\frac{ \sum_{h} {a}_{lh} {b}_{hm}}{ \|{\bm A \bm B}\|_{\rm FRO} } \right\}^{2}.\label{errVMC} 
\end{eqnarray} 
}
\end{enumerate}

In our experiment, the parameters are chosen as
$\{L, M, H, \sigma^2, \delta\}$ $= \{50, 100, 5, 0.1, 10^{-19} \}$
and covariance matrix $\bm C_{\bm A}$ is set to be identity.
We take arithmetic average of all errors over 100 experiments. 
For comparison, the numerical result of MF by
normal Gaussian prior for $\bm B$,
\begin{eqnarray}
P({\bm B}) 
& \propto & \prod_{h,m} \exp\Bigr( - \frac{ b_{hm}^2  }{2} \Bigr),
\end{eqnarray}
whose case is analyzed by VB in \cite{NS,NSBT}, is also shown.

After introducing some auxiliary variables,
the computational cost of the iteration in this experiment
is $O(LMH^2)$, 
which is mainly from the estimation of $\bar{a}_{lh}, \bar{b}_{hm}$. 
We should notice that computation of inverse matrices
$\hat{\Sigma}_{\bm A l}^{-1}, \hat{\Sigma}_{\bm B m}^{-1}$ is necessary 
during iteration.
Their costs are $O(LH^3)$ and $O(MH^3)$, which are not dominant here because
we deal with the case of small $H$, i.e. low-rank problem.

\subsection{Result 1: dependence on Laplace prior parameter $k$}
\label{result1}

\begin{figure}
\begin{picture}(0,550)
\put(0,360){\includegraphics[width=0.5\textwidth]{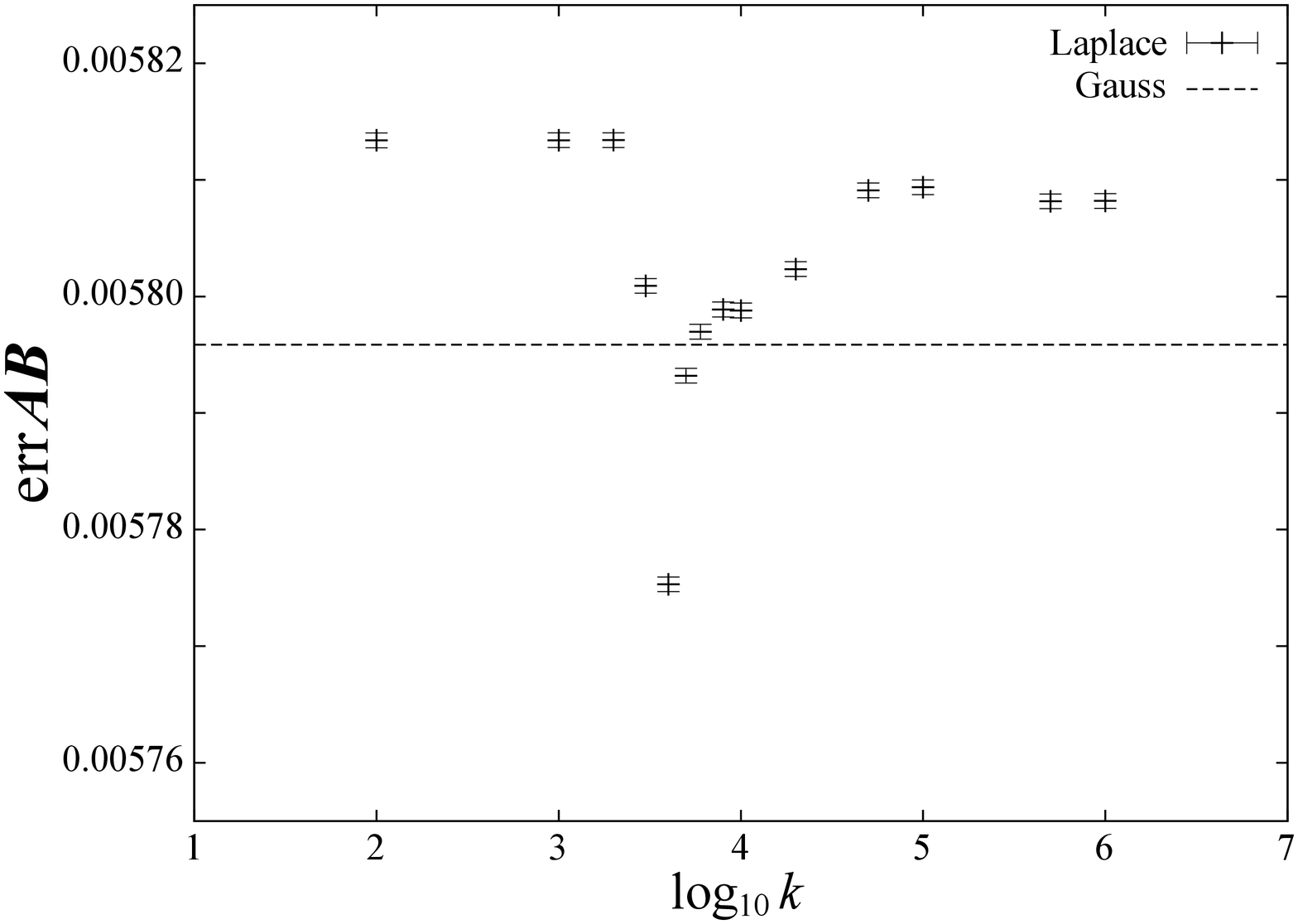}}
\put(230,359){\includegraphics[width=0.49\textwidth]{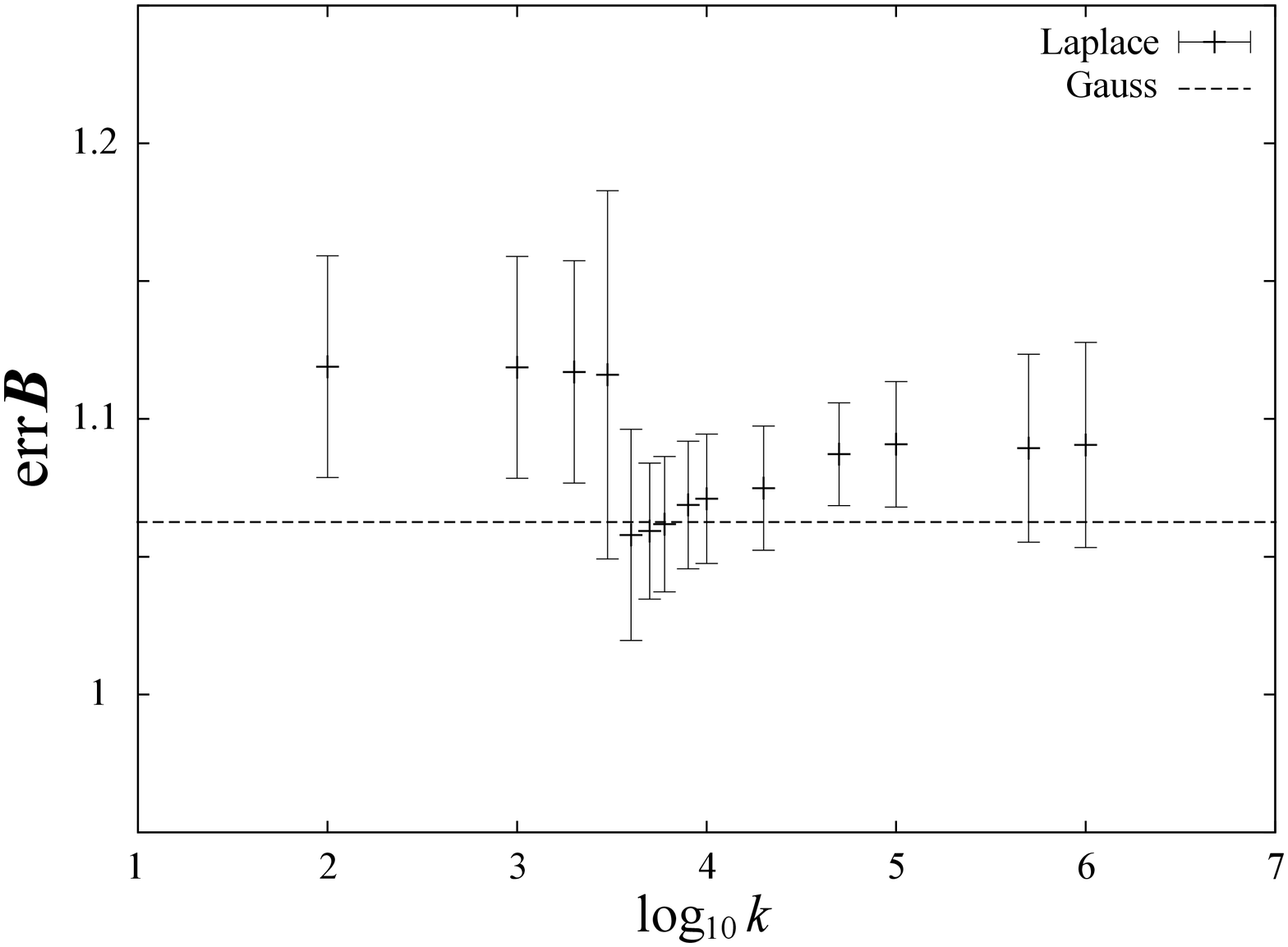}}
\put(5,180){\includegraphics[width=0.49\textwidth]{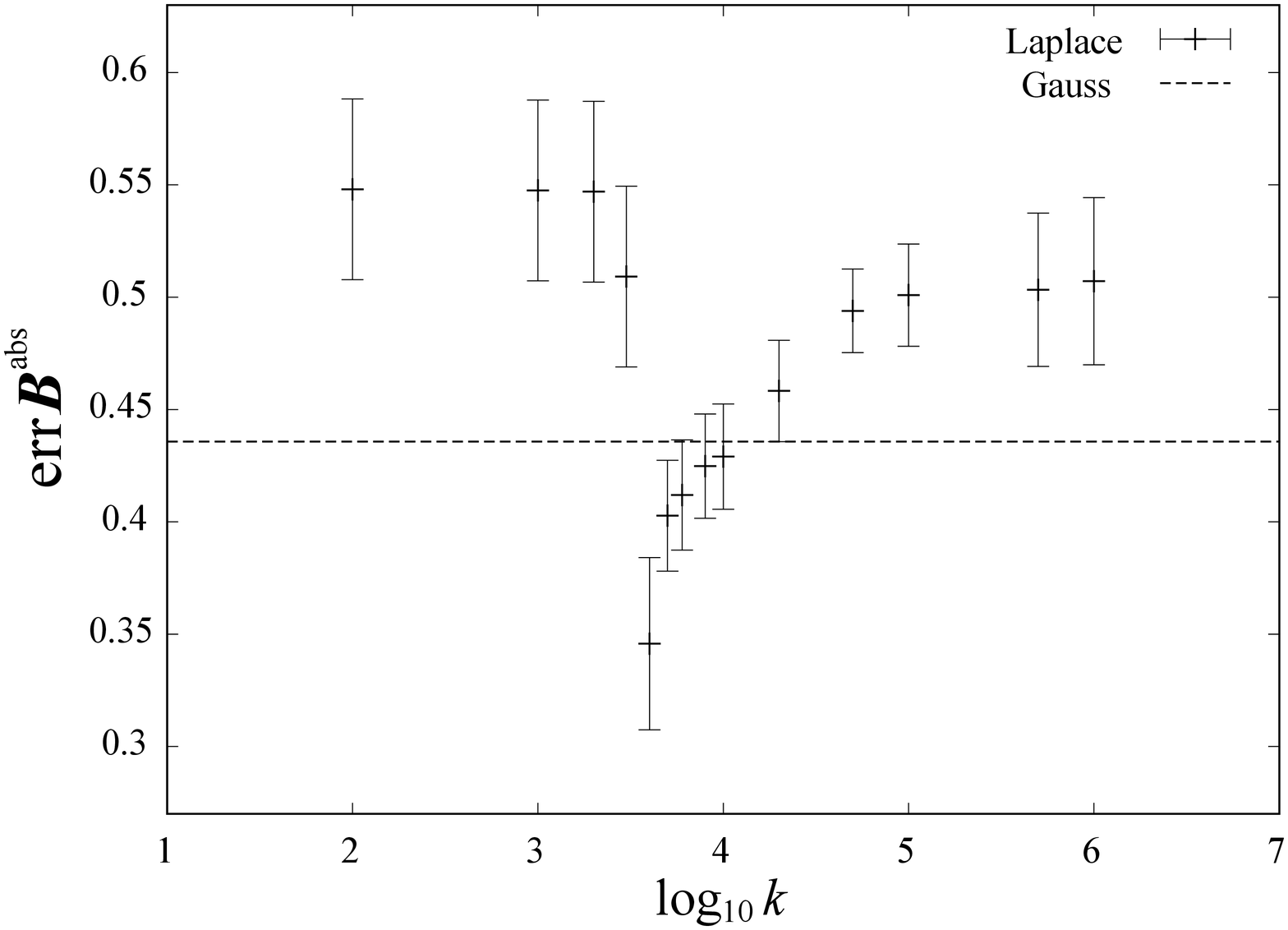}}
\put(230,182){\includegraphics[width=0.49\textwidth]{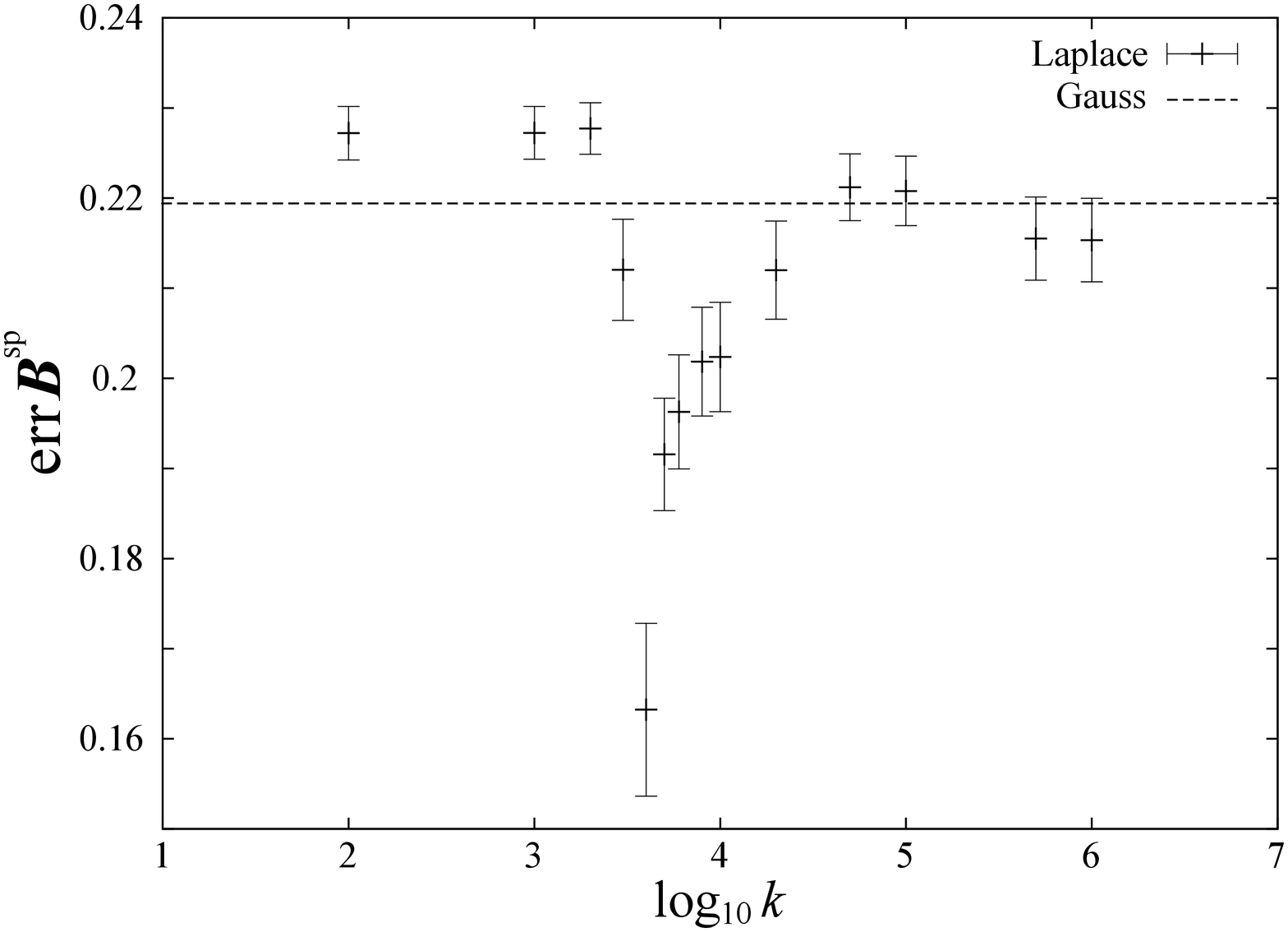}}
\put(120,2){\includegraphics[width=0.49\textwidth]{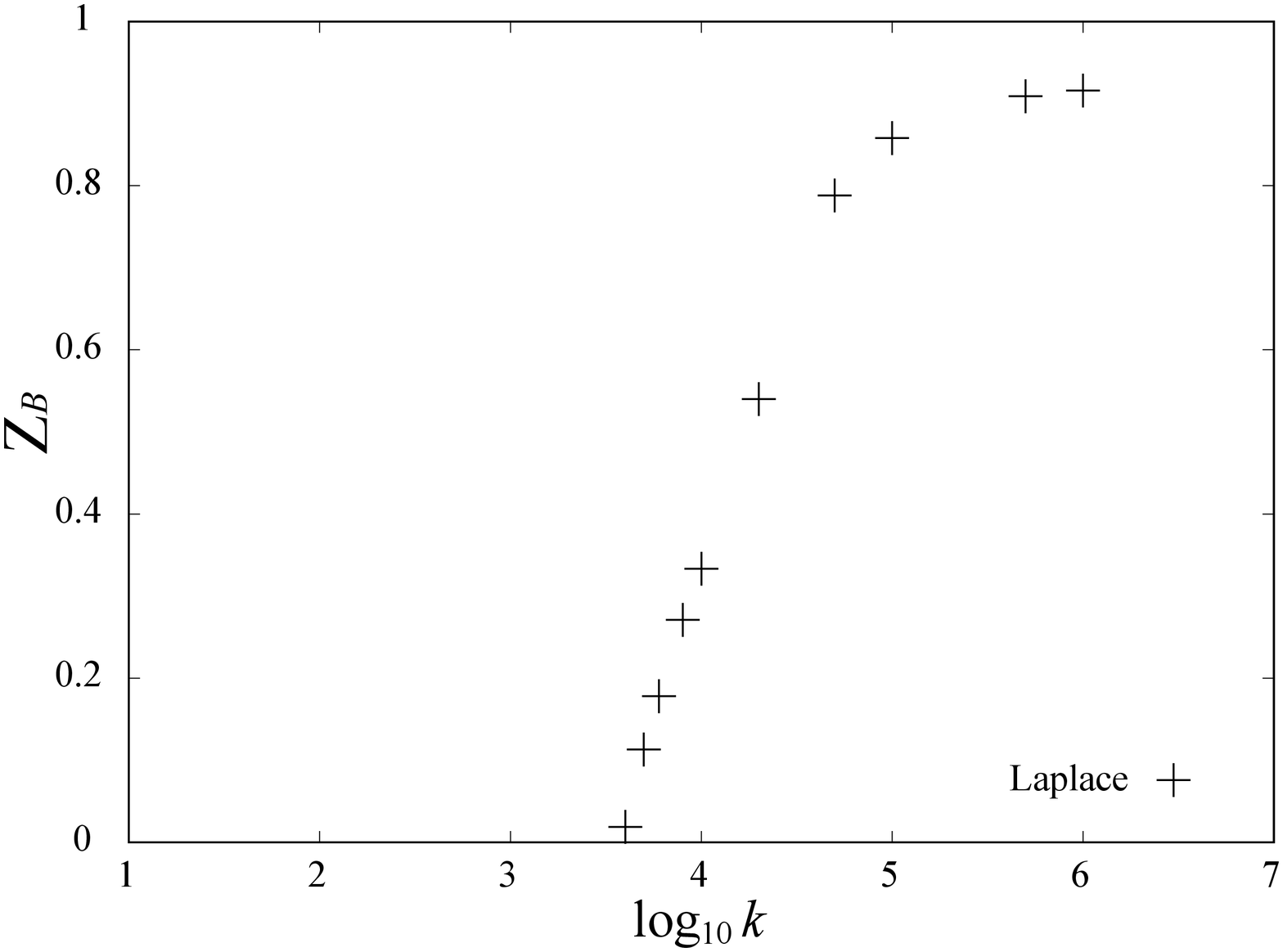}}
\end{picture}
\caption{Dependence on Laplace prior parameter $k$.}
\end{figure}
Here we vary the value of Laplace prior parameter $k$, and
conduct MF experiment for respective $k$.
The result is depicted in figure 1. 
In this experiment, the number of zero elements in $\bm B$ is 250 
(i.e. half elements in $\bm B$ are zero). 
All elements in $\bm \Theta$ is unity, namely all elements in $\bm V$ can be observed. 
We use the same $\bm \Theta$ in subsections \ref{result2} and \ref{result3} as well.

First, in the whole range of $k$, the value and the variation of ${\rm err} \bm A \bm B$ are very small. 
In other words,
Frobenius norm between the matrix product $\bm A^{(n)} \bm B^{(n)}$ after iteration 
and the original matrix product $\bm A \bm B$ is kept small and constant when $k$ is varied.
The small ${\rm err} \bm A \bm B$ guarantees the correctness of MF solution by this experiment. 
Next, the value of ${\rm err} \bm B$ is almost the same as Gaussian prior
even if we optimize $k$. On the other hand,
regarding ${\rm err} \bm B_{\rm abs}$ and ${\rm err} \bm B_{\rm sp}$
our method outperforms Gaussian prior when we choose appropriate $k$.
In particular, ${\rm err} \bm B_{\rm sp}$ by Laplace prior is much smaller than Gaussian,
which means we can identify zero elements in original $\bm B$ from observed matrix $\bm V$ more easily by Laplace prior.
We expect that $k$-dependent terms in (\ref{B1}) and (\ref{B2}), or 
corrections by Laplace prior, have significant contribution for sparse matrix reconstruction under appropriate $k$.
If $k$ is smaller than the best region, $k$-dependent denominators in (\ref{B1}) and (\ref{B2}) will
become zero or negative, which can make the mean and the variance of $\bm B$ diverge or ill-defined. 
To verify this, we also observe $k$ dependence of the renormalization factor $Z_B$ in the denominator. We easily see
that $Z_B$ becomes zero around the value of $k$ minimizing ${\rm err} \bm B_{\rm sp}$. This implies
that the best $k$ for sparse matrix reconstruction has correlation with zero point of $Z_B$.

\subsection{Result 2: dependence on fraction of zero elements in $\bm B$}
\label{result2}

Next we check the dependence on fraction of zero elements in $\bm B$, denoted by the parameter $p$ in the following.
The result is shown in figure 2.
Here we vary $p$ under fixed $k$, $k=4 \times 10^3$.
The graph of ${\rm err}{\bm B_{\rm sp}}$ shows
discrepancy between Laplace and Gaussian priors. In particular, for sparser $\bm B$,
${\rm err}{\bm B_{\rm sp}}$ by Laplace prior is much smaller, 
which indicates that Laplace prior works better for 
sparser MF.

\begin{figure}
\begin{picture}(0,190)
\put(0,0){\includegraphics[width=0.5\textwidth]{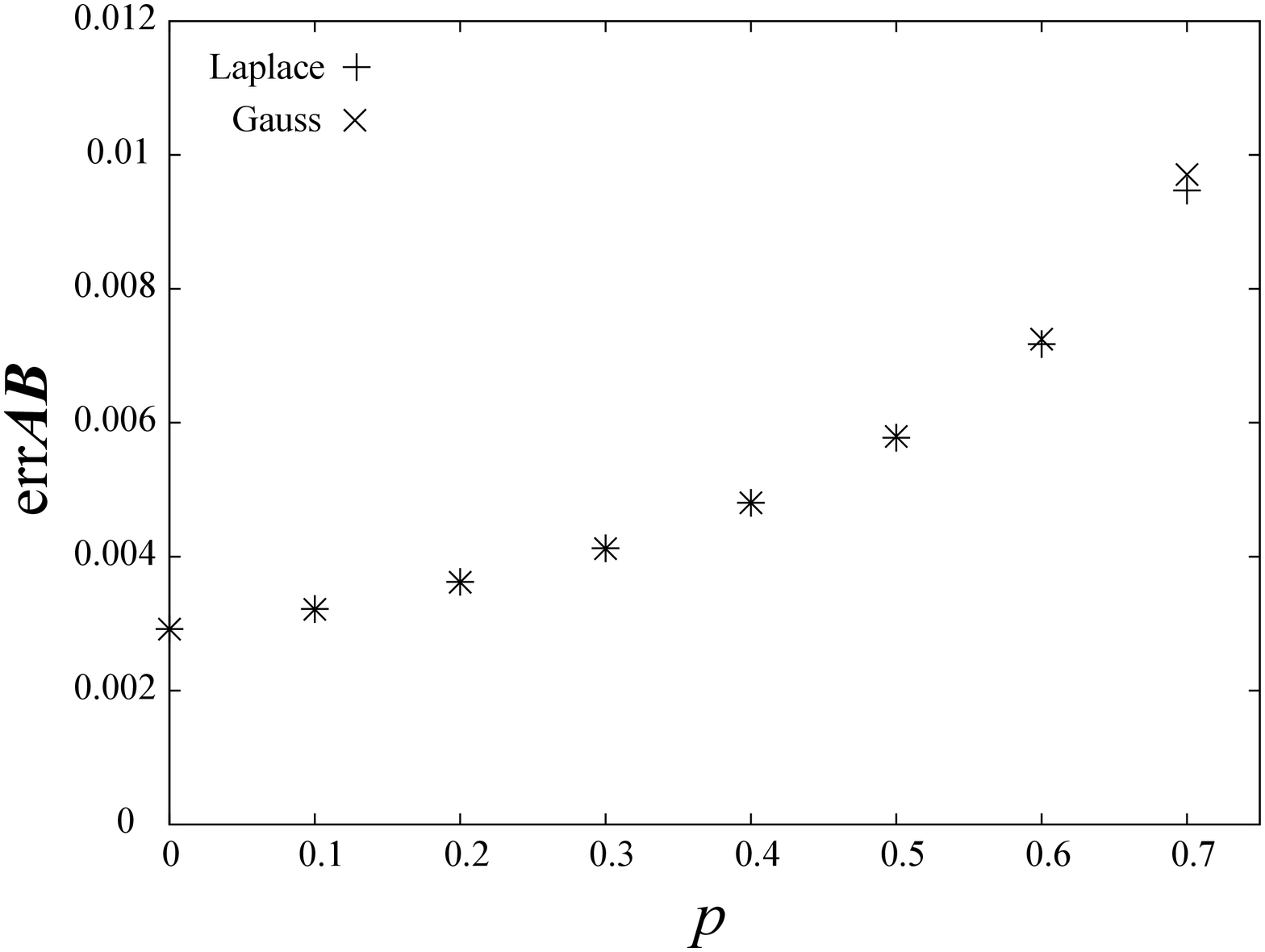}}
\put(230,2){\includegraphics[width=0.5\textwidth]{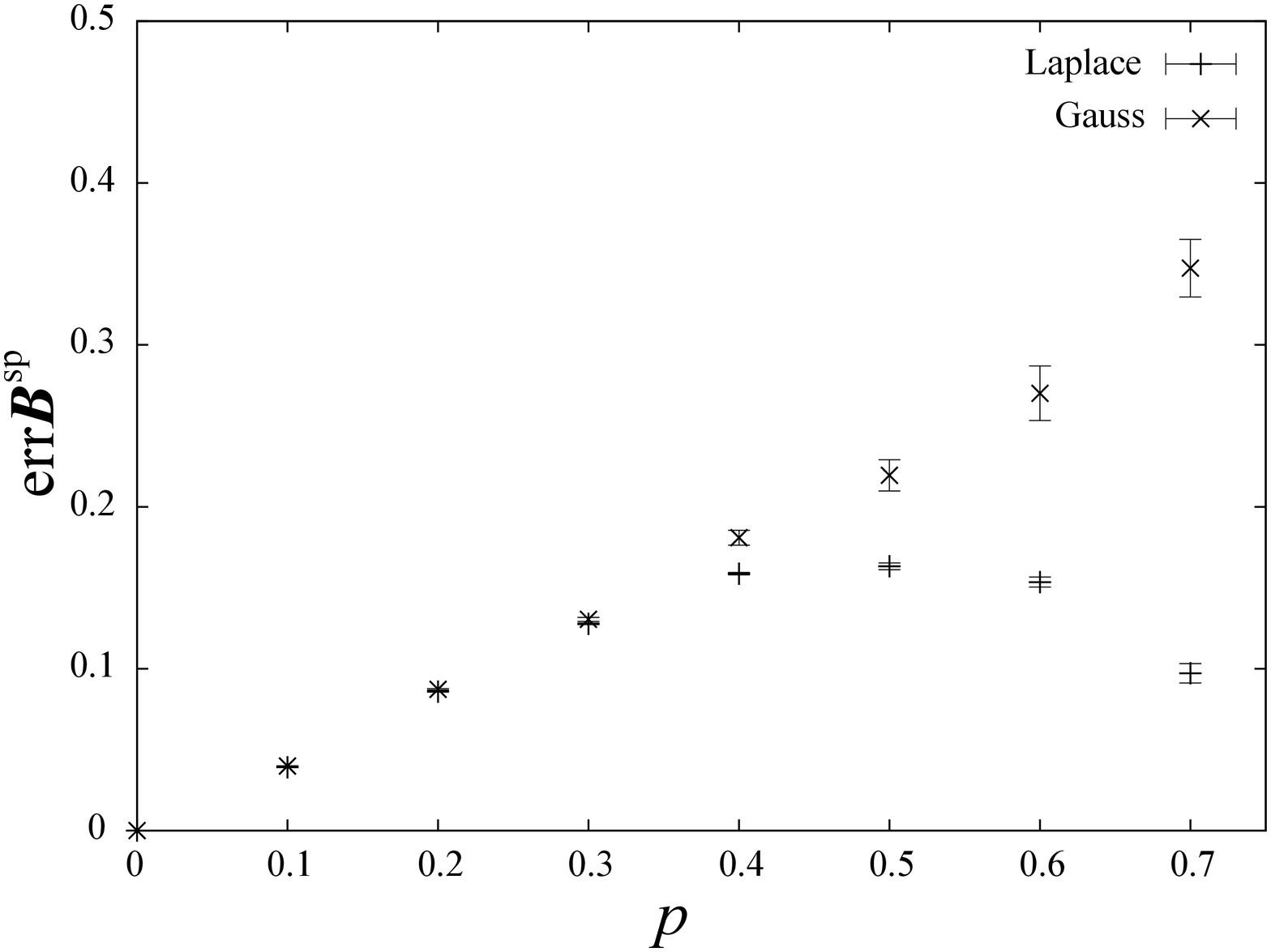}}
\end{picture}
\caption{Dependence on fraction of zero elements in $\bm B$ (denoted by $p$).
In the left figure, the error bars are very short and omitted.}
\end{figure}

It appears that the behavior of ${\rm err}{\bm B_{\rm sp}}$ by Laplace prior 
has approximate reflection symmetry with respect to the line of $p= 0.5$. 
Suppose that the numbers of zero elements in the original
$\bm B$ and the reconstructed $\bm B$ are almost the same under arbitrary $p$, and zero elements 
are randomly distributed over whole reconstructed $\bm B$. 
Then the values of ${\rm err}{\bm B_{\rm sp}}$ under $p$ and $1-p$ 
will be almost the same, because ${\rm err}{\bm B_{\rm sp}}$ only measures the error at zero elements in
original $\bm B$.
However, we do not know this symmetry is the exact one or not,
because the property of VB solution has not been revealed completely.

\subsection{Result 3: an example of MF for a given observation matrix}
\label{result3}

We directly observe matrix elements in $\bm B$ 
after MF for a given observation matrix $\bm V$.
The result is shown in figure 3.
The original $\bm B$, the reconstructed $\bm B$ by Laplace prior
after/before permutation of rows and sign adjustment,
and the one by Gaussian prior are shown by the heat maps.
Here we set $k=4 \times 10^3$ and $p=0.7$ (= 70\% of the elements in original $\bm B$ are zero) in this experiment.

\begin{figure}
\begin{picture}(0,280)
\put(10,30){\includegraphics[width=1\textwidth]{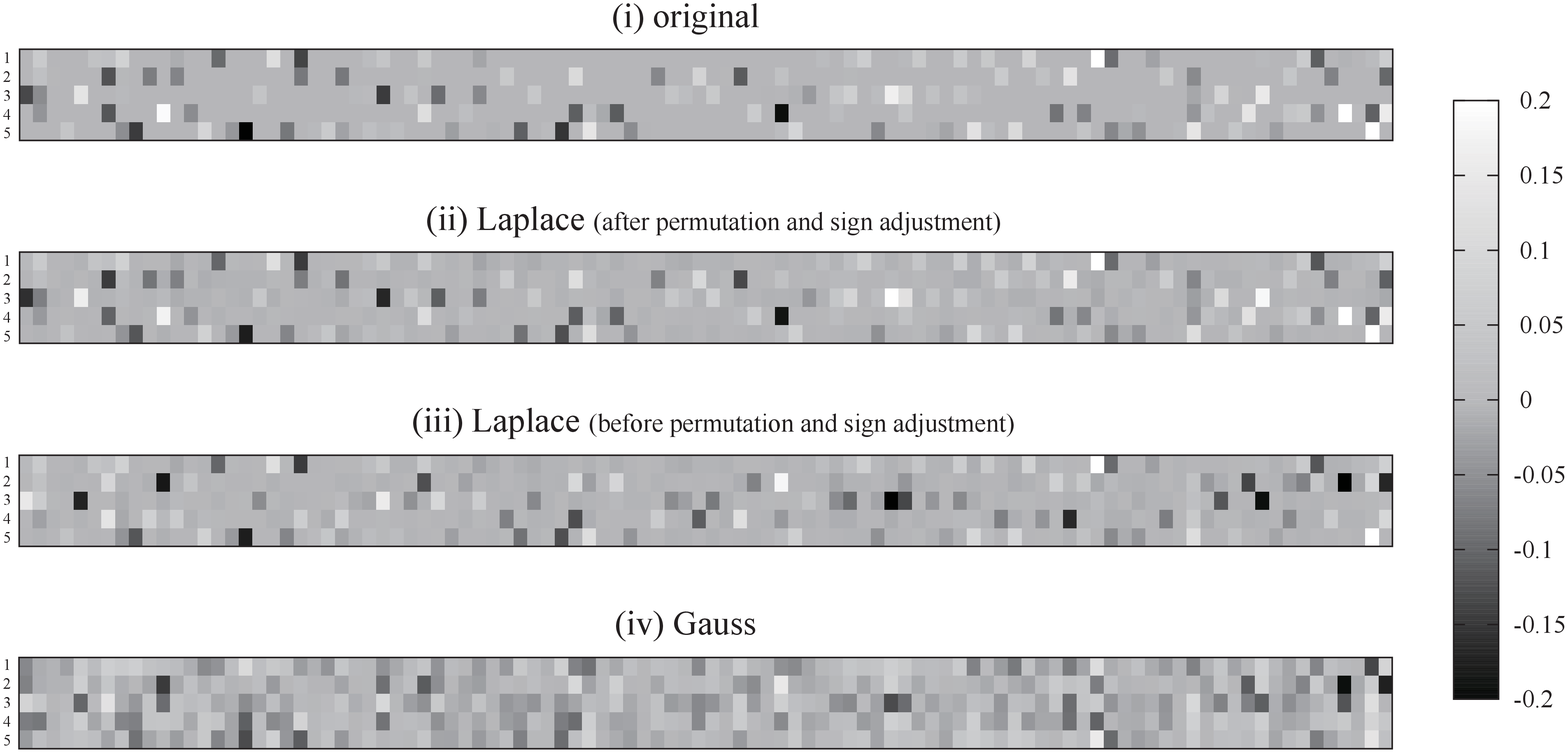}}
\end{picture}
\caption{An example of sparse matrix reconstruction. \\
From top to bottom: (i) original $\bm B$ (ii) reconstructed $\bm B$ by Laplace prior (after permutation of rows
and sign adjustment) (iii) reconstructed $\bm B$ by Laplace prior (before permutation of rows
and sign adjustment) (iv) reconstructed $\bm B$ by Gaussian prior}
\end{figure}

By comparison, we find that the reconstructed $\bm B$ by Laplace prior after 
permutation of rows and sign adjustment is almost the same as the original.
In comparing matrices, we should recall that MF solution 
has degeneracy. In general MF problem, we can obtain another MF solution
by operating rotation matrix between $\bm A$ and $\bm B$. 
Even if $\bm B$ is sparse, the matrix product $\bm A \bm B$ is invariant
 under simultaneous permutation of columns in $\bm A$ and corresponding rows in $\bm B$, 
or sign inversion of the column in $\bm A$ and the paired row in $\bm B$.
In figure 3, to obtain the most similar matrix to the original $\bm B$, 
we exchange the 2nd and the 4th rows, then reverse the signs 
of the 2nd, the 3rd, and the 4th rows.

Therefore, taking such degeneracy into consideration,
we can conclude that the original sparse matrix $\bm B$ is reconstructed practically in this example. 
In contrast, the reconstructed $\bm B$ by Gaussian prior is far from the original, which
 also supports the validity of our analysis.

\subsection{Result 4: performance of matrix completion}
We also conduct the experiment of MC.
The result is depicted in figure 4. 
In the right figure we fix $k=4 \times 10^3$.
For MC, 
$10\%$ of elements in $\bm \Theta$ are chosen to be zero. 
The performance of MC is measured by ${\rm err} \bm A \bm B^{\rm MC}$ 
with the parameter $k$ or $p$ (= fraction of zero elements in $\bm B$) being varied.
As a result, we cannot find significant difference between errors 
by Laplace/Gaussian priors. 
Although the prior affects sparse matrix reconstruction,
it will not make large contribution to the performance of MC as expected,
because the likelihood will play a central role in MC and the prior less contributes.

Similarly, at present we guess that another sparse prior like Bernoulli-Gaussian 
 will not make significant change in the result of MC.
However, we have not attempted the analysis with another prior, 
and this is the problem left for future study.

\begin{figure}
\begin{picture}(0,190)
\put(0,0){\includegraphics[width=0.5\textwidth]{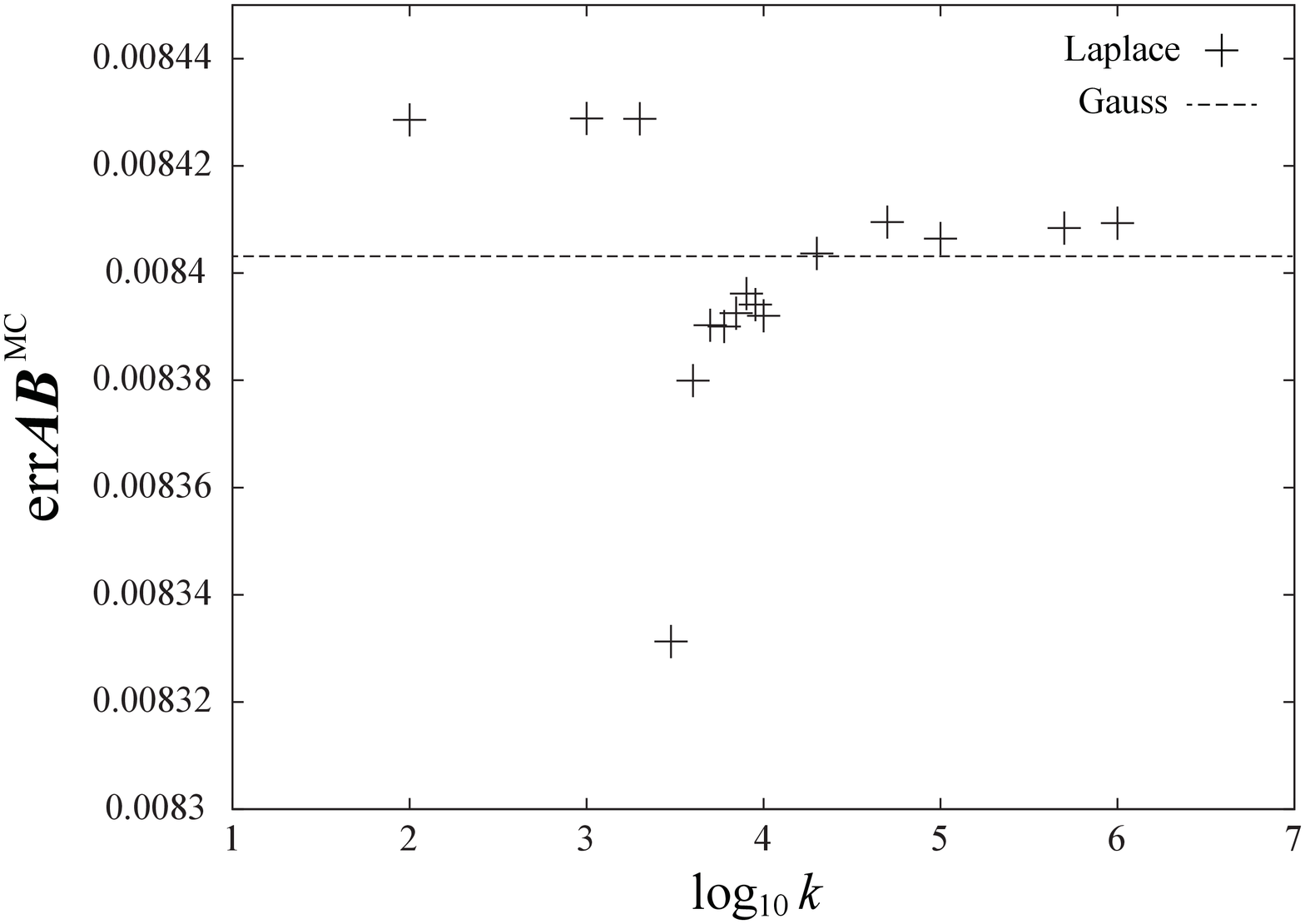}}
\put(230,1){\includegraphics[width=0.48\textwidth]{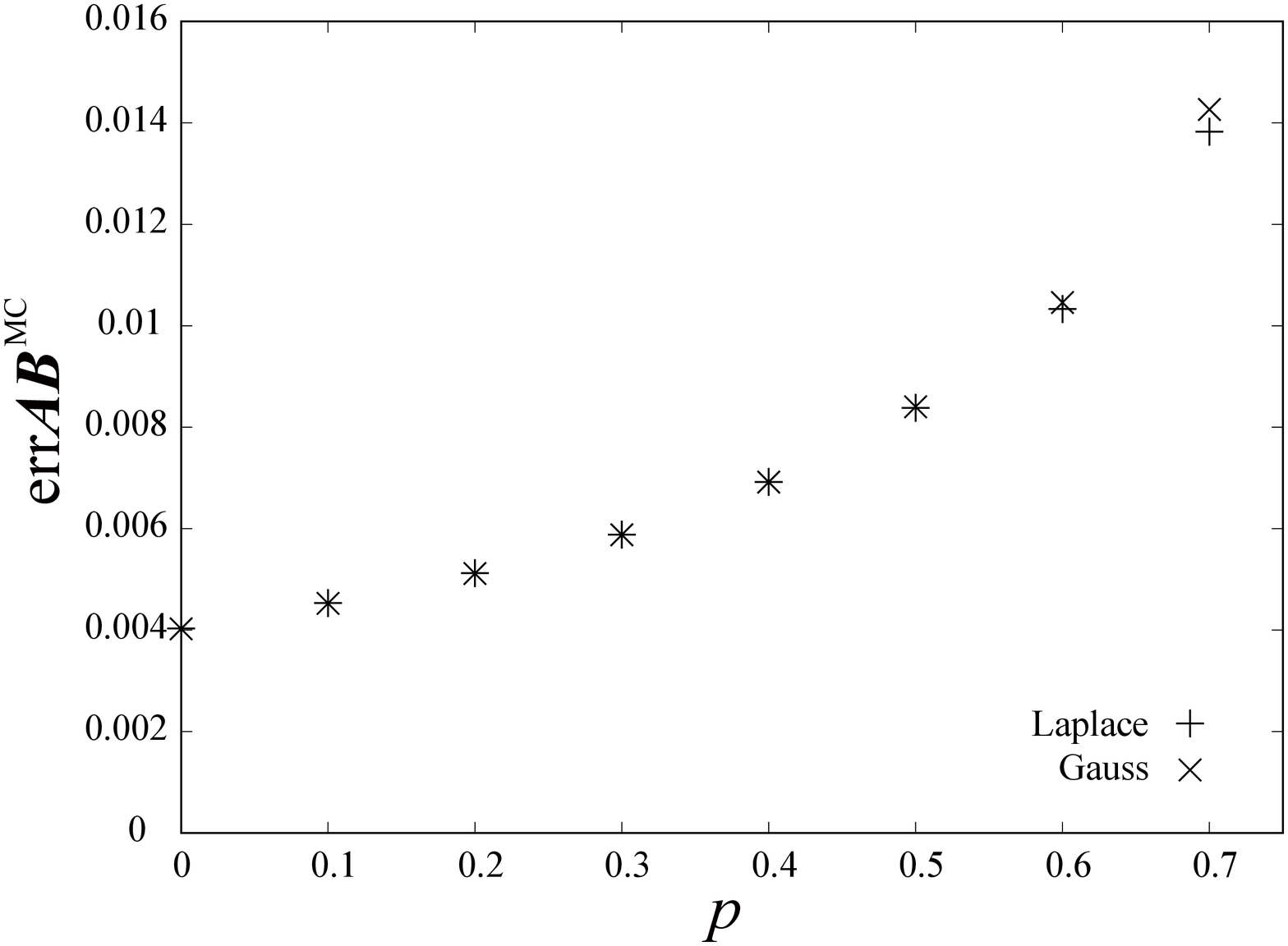}}
\end{picture}
\caption{The result of MC experiment. The error bars are omitted because they are very short and 
the difference of ${\rm err} \bm A \bm B^{\rm MC}$ between
Laplace/Gaussian priors is very small.}
\end{figure}

\section{Summary and Discussion}

We analytically obtained the solution of MF/MC by VB method under Laplace prior 
and with the first order approximation of $1/k$ expansion. 
Then we regard our solution as an iterative algorithm and conducted numerical experiment.
As a result, we found that there is a region 
of $k$ where reconstruction of sparse matrix works better than Gaussian prior. 
We also verified our method shows better performance for sparser matrix.

There remain some problems unresolved. 
First, our final equations are nonlinear, and we cannot find simpler equations between the moments 
unlike Gaussian prior. 
Due to nonlinearity, the convergence condition of the algorithm is not clear. 
The optimal $k$ giving minimum error is also a problem. As a key to this problem,
we empirically found that the best $k$ is near the zero point of the factor $Z_B$.
Then the evaluation of $Z_B$ will help determination of appropriate $k$.
Computational cost for numerical experiment is also a problem.
For reduction of the cost, we must use more efficient computational method.
Furthermore, it is difficult to guarantee the globality of KL divergence minimum unlike
Gaussian prior.

In this article, we focus on the analytical expression of the solution of MF/MC via VB.
As numerical methods, many algorithms showing good performance
have been proposed for MF/MC. We should compare the performance of MF/MC
with others as a future work. In addition, for MF/MC there are
other theoretical works based on statistical mechanical method, where
thermodynamical limit is often used. We should 
establish the relation between our method and such works.
For the present, we hope that our analysis will help the understanding of MF/MC
problem, especially dictionary learning.

Generalization of our MF analysis to another situation is also a future problem.
First, non-negative MF is significant for application, 
where all elements in matrices are non-negative.
Here we focused on real-valued matrix problem, however
it may be possible that generalization of our analysis to non-negative case 
may reveal how to address non-negative MF problem appropriately by Bayesian method.
Second, Bernoulli-Gaussian prior is also often used for describing sparsity. 
For such prior we must choose appropriate mixing ratio between delta function 
at zero and Gaussian. 
However, if we apply our formulation to Bernoulli-Gaussian prior, 
we must sum up two probabilities for delta function and Gaussian, 
which yields exponential number of factors in calculation of mean and 
covariance matrix. Therefore, another approximation must be used to deal 
with such prior.

\section*{Acknowledgement}

We are thankful to S. Nakajima for comments on related works. 
This work is supported by KAKENHI Nos. 24700007, 25120013. 

\appendix

\section{Outline of the analysis}

\subsection{Basic Gaussian case}
We start with the relation with the work in \cite{NSBT} for the sake
of clarity.
When one assumes that $P({\bm A})$ is Gaussian distribution with zero mean
and covariance matrix $\bm C_{\bm A}$ only with diagonal elements,
and $P({\bm B})$ is uniform distribution,
VB solution is obtained as in \cite{NSBT},
\begin{eqnarray}
\bar{a}_{lh} 
&=&
\sum_{m,h'}\theta_{lm} 
(\hat{ \Sigma}_{{\bm A}l}^{-1})_{hh'} v_{lm} \bar{b}_{h'm}
,\label{apA1} \\
({ \Sigma}_{{\bm A} l} )_{hh}
&=&
\sigma^2 (\hat{ \Sigma}_{{\bm A}l}^{-1} )_{hh}
,\label{apA2}
\\
\bar{b}_{hm}
&=&
\sum_{l,h'} \theta_{lm} (\hat{ \Sigma}_{{\bm B}m}^{-1})_{hh'} v_{lm} \bar{a}_{lh'}
,\label{apB1}
\\
({ \Sigma}_{{\bm B} m} )_{hh}
&=&
\sigma^2 (\hat{ \Sigma}_{{\bm B}m}^{-1})_{hh}. \label{apB2}
 \end{eqnarray}
The matrices $\hat{\bf \Sigma}_{\bm A l}, \hat{\bf \Sigma}_{\bm B m} \in {\mathbb R}^{H^2}$ are
\begin{eqnarray}
(\hat{ \Sigma}_{{\bm A}l})_{hh'}
&=&
\frac{\displaystyle \sigma^2}{\displaystyle ({ C}_{{\bm A}})_{hh}} \delta_{h,h'}
+\sum_m\theta_{lm}\Bigl(({ \Sigma}_{{\bm B} m})_{hh} \delta_{h,h'} +
\bar{b}_{hm}\bar{b}_{h'm}
\Bigr),\label{apA3}
\\
(\hat{ \Sigma}_{{\bm B} m})_{hh'}
&=&
\sum_l\theta_{lm}\Bigl(({ \Sigma}_{{\bm A }l})_{hh}\delta_{h,h'}+\bar{a}_{lh}\bar{a}_{lh'}\Bigr).
\label{apB3}
\end{eqnarray}

\subsection{$Z_B$, mean and covariance}
The limit of $k \to \infty$ is taken
for prior of $\bm B$ with keeping the terms up to $O(1/k)$,
\begin{eqnarray}
P({\bm B}) &=& \prod_{h,m}\exp \Bigr(\frac{|b_{hm}|}{k}\Bigr)
\stackrel{k \to \infty}{\longrightarrow}
1 -\sum_{h,m}\frac{|b_{hm}|}{k}.
\label{apApproxB}
\end{eqnarray}
$P(\bm A)$ is the same as in \cite{NSBT}, therefore the quantities for $\bm A$,
$\bar{a}_{lh}$ in (\ref{apA1}), $(\Sigma_{\bm A l})_{hh}$ in (\ref{apA2}), 
and $(\hat{\Sigma}_{\bm A l})_{hh'}$ in (\ref{apA3}) do not change.
One needs to calculate the quantities for $\bm B$, namely $\bar{b}_{hm},
(\Sigma_{\bm B m})_{hh}$, and $(\hat{\Sigma}_{\bm B m})_{hh'}$.
After insertion of (\ref{apApproxB}) into (\ref{r2}) and using (\ref{meanb}), 
the mean and the variance, which are finally evaluated, 
are expressed as
\begin{eqnarray}
\bar{b}_{hm}
&=&
\frac{1}{Z_B F}\Bigr\{
\int_{{\mathbb R}^{HM}} b_{hm} 
\left( 1 - \sum_{h',m'} \frac{ |b_{h'm'}| }{k} \right) f({\bm B})d{\bm B}\Bigr\},
\label{bmean}
\\
({\bf \Sigma}_{{\bm B} m})_{hh}
&=&
\frac{1}{Z_B F}
\Bigr\{
\int_{{\mathbb R}^{HM}} b_{hm}^2 
\left( 1 - \sum_{h',m'} \frac{ |b_{h'm'}| }{k} \right)
f({\bm B})d{\bm B} \Bigr\}
-\bar{b}_{hm}^2.
\label{bvariance}
\end{eqnarray}
The factor $Z_BF$ serves as partition function for evaluation of the moments.
$Z_B$ is the renormalization factor of Laplace prior,
because one must renormalize the probability distribution of $\bm B$
due to the approximation up to $O(1/k)$. 
The zero point of $Z_B$ describes the boundary between well- and ill-defined regions of 
renormalized probability.
\begin{eqnarray}
Z_B &:=&  \frac{1}{F}
\int_{{\mathbb R}^{HM}} \left( 1 - \sum_{h,m} \frac{ |b_{hm}| }{k} \right) f({\bm B})d{\bm B}, \\ 
F &:=& \int_{{\mathbb R}^{HM}} f({\bm B})d{\bm B}
\end{eqnarray}
where $f(B)$ is the exponential weight,
\begin{eqnarray}
f({\bm B})
&:=&
\prod_{h,m}
\exp
\Bigr\{
 -
\frac{1}{2\sigma^2}
\Bigr(
\alpha_{hhm}
b_{hm}^2 
+ 
\sum_{h' \ne h}
\alpha_{hh'm}
b_{hm}b_{h'm}
-2
\gamma_{hm}
b_{hm}
\Bigr)\Bigr\}, \nn\\
\end{eqnarray}
and one defines the variables,
\begin{eqnarray}
\alpha_{hh'm}
&:=&
\sum_l \theta_{lm}\Bigr(( \Sigma_{{\bm A}l})_{hh}\delta_{h,h'} +\bar{a}_{lh}\bar{a}_{lh'}\Bigr),
\\
\gamma_{hm}&:=&
\sum_l \theta_{lm} v_{lm}\bar{a}_{lh}.\label{defgamma}
\end{eqnarray}

\subsection{Integration of $f({\bm B})$}
For performing the multiple integrals
$\int_{{\mathbb R}^{HM}} f({\bm B})d{\bm B}$, one defines the matrices and
vectors as
\begin{eqnarray}
{\bm b}^{T}
&:=& ( {\bm b}_1^T, {\bm b}_2^T, \ldots \ {\bm b}_M^T ), \\
{\bm b}_m^T
&:=& ( b_{1m}, b_{2m}, \ldots, b_{Hm} ), \\
{\bm \gamma }^T
&:=& ( {\bm \gamma}_1^T, {\bm \gamma}_2^T, \ldots {\bm \gamma}_M^T ), \\
{\bm \gamma}_m^T
&:=& (\gamma_{1m}, \gamma_{2m}, \ldots \gamma_{Hm} ), \\
{\hat{\bm \Sigma}_{\bm B}}
&:=&
\left(\begin{array}{cccc}
\hat{\bm \Sigma}_{{\bm B} 1} & &  & \\
 & \hat{\bm \Sigma}_{{\bm B} 2} &&  $\mbox{\large 0}$\\
& &\ddots &\\
 $\mbox{\large 0}$ &  & & \hat{\bm \Sigma}_{{\bm B} M}
\end{array}  \right),
\ {\rm where} \ \
 (\hat{ \Sigma}_{{\bm B} m})_{ij}
:= \alpha_{i j m}.
\end{eqnarray}
By using them, one calculates the integral
\begin{eqnarray}
F&:=&\int_{{\mathbb R}^{HM}} f({\bm B})d{\bm B} \nn \\
&=&
\int_{{\mathbb R}^{HM}}
\exp
\Bigr\{-\frac{1}{2\sigma^2}
\Bigr(
{\bm b}^T{\hat{\bm \Sigma}_B}{\bm b} -2{\bm \gamma}^T{\bm b}
\Bigr)\Bigr\}d{\bm B}
\nn\\
&=&
(2\pi\sigma^2)^{\frac{1}{2}HM}\prod_{m} (\det \hat{\bm \Sigma}_{{\bm B} m})^{-\frac{1}{2}}
\exp\Bigr( \frac{1}{2\sigma^2}
{\bm \gamma}_{m}^T \hat{\bm \Sigma}_{{\bm B} m}^{-1}{\bm \gamma}_{m}   \Bigr).
\end{eqnarray}
Next one evaluates the integral 
\begin{eqnarray}
\tilde{F}_{hm} &:=& \int_{{\mathbb R}^{HM -1}} f({\bm B})d({\bm B}\backslash b_{hm}),
\label{fBintegral-before}
\end{eqnarray}
where ${\bm B}\backslash b_{hm}$ represents all elements in $\bm B$ excepting $b_{hm}$.
As a result,
\begin{eqnarray}
\tilde{F}_{hm} &=&
(2\pi\sigma^2)^{\frac{1}{2} HM - \frac{1}{2}H} \prod_{m' \neq m} (\det \hat{\bm \Sigma}_{{\bm B} m'})^{-\frac{1}{2}}
\exp\Bigr( \frac{1}{2\sigma^2}
{\bm \gamma}_{m'}^T \hat{\bm \Sigma}_{{\bm B} m'}^{-1} {\bm \gamma}_{m'} \Bigr)
\nn\\
&&\hspace{0cm}
\times
\int_{{\mathbb R}^{H-1}}
\exp
\Bigr\{-\frac{1}{2\sigma^2}
\Bigr(
(\hat{ \Sigma}_{{\bm B} m})_{hh} b_{hm}^2 -2\gamma_{hm}b_{hm}
\nn\\
&&\hspace{2.5cm}
+
({\bm b}_{m\backslash h})^T(\hat{\bm \Sigma}_{{\bm B} {m\backslash h}})
({\bm b}_{m\backslash h}) -2{\bm \tau}_{mh}^T({\bm b}_{m\backslash h})
\Bigr)\Bigr\}
d {\bm B}_{h \backslash m}
\nn\\
&=&
(2\pi\sigma^2)^{\frac{1}{2}HM-\frac{1}{2}}
(\det \hat{\bm \Sigma}_{{\bm B} {m \backslash h}})^{-\frac{1}{2}}
\nn\\
&& \times \prod_{m' \neq m} (\det \hat{\bm \Sigma}_{{\bm B} m'})^{-\frac{1}{2}}
\exp\Bigr( \frac{1}{2\sigma^2}
{\bm \gamma}_{m'}^T \hat{\bm \Sigma}_{{\bm B} m'}^{-1} {\bm \gamma}_{m'}   \Bigr)
\nn\\
&&\hspace{0cm}
\times
\exp
\Bigr\{-\frac{1}{2\sigma^2}
\Bigr(
(\hat{ \Sigma}_{\bm B m})_{hh} b_{hm}^2 -2\gamma_{hm}b_{hm}
+
{\bm \tau}_{mh}^T (\hat{\bm \Sigma}_{\bm B m\backslash h}^{-1})_* {\bm \tau}_{mh} 
\Bigr)\Bigr\}. \nn \\
\label{fBintegral}
\end{eqnarray}
The vector ${\bm \tau}_{mh}$ is defined by $({ \tau}_{mh})_{h'} := 
( \gamma_m)_{h'} - (\hat{ \Sigma}_{\bm B m})_{h'h} ( b_m)_h =
\gamma_{h'm} - \alpha_{h'hm} b_{hm} \label{tau}$, and
${\bm b}_{m \backslash h}\ \in \mathbb{R}^{H-1}$ is given by
removing $h$-th element from $\bm b_m$.
Similarly, the matrix $\hat{\bm \Sigma}_{\bm B m \backslash h} \in \mathbb{R}^{(H-1)^2}$ is 
given by removing $h$-th column and row from $\hat{\bm \Sigma}_{\bm B m}$.
The matrix $(\hat{\bm \Sigma}_{\bm B m \backslash h})^{-1}_* \in \mathbb{R}^{H^2}$
is defined by inserting zero vector in 
$h$-th column and row in $(\hat{\bm \Sigma}_{\bm B m \backslash h}^{-1})$, 
namely
\begin{eqnarray}
(\hat{ \Sigma}_{\bm B m \backslash h}^{-1})_{* h'h''} :=
\left\{
\begin{array}{ll}
(\hat{ \Sigma}_{\bm B m \backslash h}^{-1})_{h'h''} & h'<h, h''<h,\\
(\hat{ \Sigma}_{\bm B m \backslash h}^{-1})_{h'-1,h''} & h'>h, h''<h,\\
(\hat{ \Sigma}_{\bm B m \backslash h}^{-1})_{h',h''-1} & h'<h, h''>h,\\
(\hat{ \Sigma}_{\bm B m \backslash h}^{-1})_{h'-1,h''-1} & h'>h, h''>h,\\
0 & h'=h\ \ {\rm or}\ \ h''=h.
\end{array}
\right.
\end{eqnarray}
The factor ${\bm \tau}_{mh} ^T (\hat{\bm \Sigma}_{\bm B m\backslash h}^{-1})_*{\bm \tau}_{mh} $
is separated into $O(b_{hm}^2)$, $O(b_{hm})$, and $O(1)$ terms.
\begin{eqnarray}
&&{\bm \tau}_{mh}^T (\hat{\bm \Sigma}_{\bm B m\backslash h}^{-1} )_* {\bm \tau}_{mh}
\nn \\
&=&
\sum_{h',h''}
(\hat{\Sigma}_{\bm B m\backslash h}^{-1})_{*h'h''} 
\nn \\ 
&& \times
\{ 
(\hat{ \Sigma}_{\bm B m})_{h'h} (\hat{ \Sigma}_{\bm B m})_{h''h} b_{hm}^2 \nonumber \\
&&
\ \ \ \ - ( (\hat{ \Sigma}_{\bm B m})_{h'h} \gamma_{h'' m} + (\hat{ \Sigma}_{\bm B m})_{h''h} \gamma_{h' m}) b_{hm}
+ \gamma_{h' m} \gamma_{h'' m} \}. 
\end{eqnarray}
Here one defines $x_{hm}, y_{hm}$ and $z_{hm}$ as the coefficients of 
$O(b_{hm}^2)$, $O(b_{hm})$, and $O(1)$ terms in (\ref{fBintegral}).
\begin{eqnarray}
&&
 \exp \Bigr\{ -\frac{1}{2 \sigma^2} \Bigr(
(\hat{ \Sigma}_{\bm B m})_{hh} b_{hm}^2 -2\gamma_{hm}b_{hm}
+
{\bm \tau}_{mh} ^T (\hat{\bm \Sigma}_{\bm B m\backslash h}^{-1} )_*{\bm \tau}_{mh}
\Bigr) \Bigr\}
\nn\\
&=:&
\exp \Bigr\{ -\frac{1}{2 \sigma^2} \Bigr(
x_{hm} b_{hm}^2  - 2y_{hm}b_{hm} - z_{hm} \Bigr) \Bigr\}.
\end{eqnarray}
From comparison of $O(b_{hm}^2)$, $O(b_{hm})$, and $O(1)$
terms, one obtains
\begin{eqnarray}
x_{hm}
&:=&
(\hat{ \Sigma}_{\bm B m})_{hh}
-
\sum_{h',h''} 
(\hat{ \Sigma}_{\bm B m \backslash h}^{-1})_{*h'h''}
(\hat{ \Sigma}_{\bm B m})_{h'h}
(\hat{ \Sigma}_{\bm B m})_{h''h}
=
\frac{1}{(\hat{ \Sigma}_{\bm B m}^{-1} )_{hh}},
\label{x}
\nn \\
 \\
y_{hm}
&:=&
\gamma_{hm}
-
\frac{1}{2}
\sum_{h',h''} 
(\hat{ \Sigma}_{\bm B m\backslash h}^{-1})_{*h'h''}
((\hat{ \Sigma}_{\bm B m})_{h'h}
\gamma_{h'' m} 
+ (\hat{ \Sigma}_{\bm B m})_{h''h}  \gamma_{h' m}  )
\nn \\
&=&
\frac{
\sum_{h'} (\hat{ \Sigma}_{\bm B m}^{-1})_{hh'} \gamma_{h'm}
}{(\hat{ \Sigma}_{\bm B m}^{-1})_{hh}},
\label{y}
\\
z_{hm}
&:=&
\sum_{h',h''} 
(\hat{ \Sigma}_{\bm B m\backslash h}^{-1})_{*h',h''}
\gamma_{h' m} \gamma_{h'' m}  
=
({\bm \gamma}_{m})^T
 (\hat{\bm \Sigma}_{\bm B m\backslash h}^{-1})_* 
({\bm \gamma}_{m}).
\label{z}
\end{eqnarray}
The expression of $x_{hm}$ is given by Schur complement for $\hat{\bm \Sigma}_{\bm B m}$.
The expression of $y_{hm}$ can be obtained by the matrix $\hat{\bm \Sigma}_{\bm B m}^{(\gamma h)}$, which is defined
by replacing $h$-th column or row in $\hat{\bm \Sigma}_{\bm B m}$ with the vector $\bm \gamma$,
and by considering Schur complement for $\hat{\bm \Sigma}_{\bm B m}^{(\gamma h)}$ in conjunction with
the identity $(\hat{ \Sigma}_{\bm B m}^{-1})_{hh} =
(\hat{ \Sigma}_{\bm B m}^{(\gamma h) -1})_{hh}  \sum_{h'} (\hat{ \Sigma}_{\bm B h}^{-1})_{hh'} \gamma_{h'm}$.

Then $\tilde{F}_{hm}$ can be rewritten as
\begin{eqnarray}
\tilde{F}_{hm} 
&=& (2\pi\sigma^2)^{\frac{1}{2}HM - \frac{1}{2}}
(\det \hat{\bm \Sigma}_{\bm B m\backslash h})^{-\frac{1}{2}}
\prod_{m' \ne m} (\det \hat{\bm \Sigma}_{\bm B m'})^{-\frac{1}{2}}
\nn \\
&& \!\!\!\! \times 
\exp\Bigr\{
-\frac{1}{2\sigma^2}
\Bigr(
x_{hm} b_{hm}^2 - 2y_{hm}b_{hm} -z_{hm}
-
\sum_{m' \neq m}
{\bm \gamma}_{m'}^T (\hat{\bm \Sigma}_{\bm B m'}^{-1}){\bm \gamma}_{m'}  
\Bigr)\Bigr\}. \nn \\
\end{eqnarray}
One should note that $F= \int_{\mathbb R} \tilde{F}_{hm}  db_{hm} 
=
\int_{\mathbb R} \Bigr\{ \int_{\mathbb R^{HM-1}} f({\bm B})d({\bm B}\backslash b_{hm}) \Bigr\} db_{hm} $.
\label{finalZB}
\subsection{Evaluation of $Z_B$}

Before evaluation of $Z_B$, it is convenient to calculate
the integration in the following,
\begin{eqnarray}
&& \int_{\mathbb R} {\rm sign} (b_{h m}) \tilde{F}_{hm} db_{h m}
\nn \\
&=&  (2\pi\sigma^2)^{\frac{1}{2}HM - \frac{1}{2}}
(\det \hat{\bm \Sigma}_{\bm B m \backslash h})^{-\frac{1}{2}}
\prod_{m' \ne m} (\det \hat{\bm \Sigma}_{\bm B m'})^{-\frac{1}{2}}
\nn \\
&& \times
\int_{\mathbb R} {\rm sign} (b_{h m})
\exp\Bigl\{
-\frac{1}{2\sigma^2}
\Bigr(
x_{h m}b_{h m}^2 -2 y_{h m}b_{h m} - z_{h m} \nn \\
&& \hspace{7cm} 
 - \sum_{m'} {\bm \gamma}_{m'}^T (\hat{\bm \Sigma}_{\bm B m'}^{-1}){\bm \gamma}_{m'} 
\Bigr)\Bigr\}
db_{h m}
\nn \\
&=&
 \Bigr( \frac{x_m}{2 \pi \sigma^2} \Bigr)^{\frac{1}{2}}
\Bigr[
\int_0^{\infty} \exp\Bigr\{ -\frac{ x_{h m} }{2\sigma^2}
\Bigr(b_{h m} -\frac{y_{h m}}{x_{h m}} \Bigr)^2 \Bigr\}
db_{h m} \nn \\
&&
\hspace{3cm} +
\int_0^{-\infty} \exp\Bigr\{ -\frac{ x_{h m} }{2\sigma^2}
\Bigr(b_{h m} -\frac{y_{h m}}{x_{h m}} \Bigr)^2 \Bigr\}
db_{h m}
\Bigr]F
\nn\\
&=&
\Bigr( \frac{1}{2 \pi} \Bigr)^{\frac{1}{2}}
\Bigr\{\int_{-\frac{y_{h m}}{\sqrt{2\sigma^2 x_{h m}}}}^\infty 
\exp(-t_{h m}^2)dt_{h m}
+ 
\int_{-\frac{y_{h m}}{\sqrt{2\sigma^2 x_{h m}}}}^{-\infty}    
\exp(-t_{h m}^2)dt_{h m}\Bigr\}F
\nn\\
&=&
{\rm erf}\Bigr(  \frac{y_{h m}}{\sqrt{2\sigma^2 x_{h m}}}  \Bigr)F,
\end{eqnarray}
from which one defines $\epsilon_{hm}$,
\begin{eqnarray}
&&
\sigma^2\frac{\partial }{\partial \gamma_{h m}}
\int_{\mathbb R} {\rm sign} (b_{h m}) \tilde{F}_{hm} db_{h m}
\nn\\
&=&
\sigma^2\frac{\partial }{\partial \gamma_{h m}}
\Bigr\{ 
{\rm erf}\Bigr(  \frac{y_{h m}}{\sqrt{2\sigma^2 x_{h m}}}  \Bigr)F \Bigr\}
\nn\\
&=&
\Biggr\{
\sqrt{\frac{2 \sigma^2}{\pi x_{h m}}}
\exp\Bigr(-\frac{y_{h m}^2}{2\sigma^2x_{h m}}\Bigr)
+
\Bigr( \sum_{h'} (\hat{\bm \Sigma}_{\bm B m}^{-1})_{hh'} \gamma_{h'm} \Bigr) 
{\rm erf}\Bigr(  \frac{y_{h m}}{\sqrt{2\sigma^2 x_{h m}}}  \Bigr)
\Biggr\} F
\nn\\
&=: &
\epsilon_{h m} F.
\end{eqnarray}
Then $Z_B$ is expressed as
\begin{eqnarray}
Z_B
&=& \frac{1}{F} \left\{
\int_{{{\mathbb R}^{HM}}} f({\bm B})d{\bm B}
-
\frac{1}{k} \sum_{h ,m} \int_{{\mathbb R}^{HM}}
 |b_{h m}| f({\bm B})d{\bm B} \right\}
\nn\\
&=&
\frac{1}{F} \left\{ \int_{{\mathbb R}^{HM}} f({\bm B})d{\bm B}
-
\frac{1}{k} \sum_{h,m}  \int_{\mathbb R} {\rm sign}(b_{hm}) b_{hm} \tilde{F}_{h m} db_{hm}
\right\} \nn\\
&=&
1
-
\frac{1}{kF} \sum_{h,m}  \sigma^2\frac{  \partial }{\partial \gamma_{hm}} \int_{\mathbb R}
{\rm sign}(b_{hm}) \tilde{F}_{h m} db_{h m}
\nn\\
&=&
 1 - \frac{1}{k} \sum_{h, m} \epsilon_{h m}.
\label{finalZB}
\end{eqnarray}

\subsection{Evaluation of mean}

Before evaluation of mean, one calculates
\begin{eqnarray}
&& \sigma^4 \frac{\partial}{\partial \gamma_{hm} }
\Bigr(
\frac{\partial }{\partial \gamma_{h' m'}}
\int_{{\mathbb R}} {\rm sign}(b_{h' m'}) \tilde{F}_{h' m'} db_{h' m'}
\Bigr) \nn \\
&=& \sigma^2 \frac{\displaystyle \partial}{\displaystyle \partial \gamma_{hm} }
\Bigr(\epsilon_{h' m'} F \Bigr)
= 
\sigma^2
\Bigr(
\frac{\displaystyle \partial \epsilon_{h' m'}  }{\displaystyle \partial \gamma_{h m} }  F
+ \epsilon_{h' m'}
\frac{\displaystyle \partial F}{\displaystyle \partial \gamma_{h m} }
\Bigr) 
\nn \\
&=&
\frac{\sigma^2}{x_{h' m}}
{\rm erf}\Bigr(\frac{y_{h' m}}{\sqrt{2\sigma^2 x_{h' m}}} \Bigr) 
\frac{(\hat{ \Sigma}_{\bm B m}^{-1} )_{h' h} }{(\hat{ \Sigma}_{\bm B m}^{-1} )_{h' h'}} 
\delta_{m,m'} F
+
\epsilon_{h' m'}
\Bigr( \sum_{m''} (\hat{ \Sigma}_{\bm B m}^{-1} )_{hh}\gamma_{hm''} \Bigr) F.
\nn \\
\label{A5-first}
\end{eqnarray}
In the last line of (\ref{A5-first}), the factor below is calculated as
\begin{eqnarray}
&& \sigma^2
\frac{\partial \epsilon_{h' m'}  }{\partial \gamma_{h m} }  
\nn \\
&=&
\sigma^2
\frac{\partial }{\partial \gamma_{h m} }
\Biggr\{
\sqrt{\frac{2 \sigma^2}{\pi x_{h' m'}}}
\exp\Bigr(-\frac{y_{h' m'}^2}{2\sigma^2x_{h' m'}}\Bigr)
+
\frac{y_{h' m'}}{x_{h' m'}} 
{\rm erf}\Bigr(  \frac{y_{h' m'}}{\sqrt{2\sigma^2 x_{h' m'}}}  \Bigr)
\Biggr\} 
\nn \\
&=&
\Biggr\{-
\sqrt{\frac{2 \sigma^2}{\pi x_{h' m'}}}
\frac{y_{h' m'}}{x_{h' m'}} 
\exp\Bigr(-\frac{y_{h' m'}^2}{2\sigma^2x_{h' m'}}\Bigr)
+
\frac{\sigma^2}{x_{h' m'}} {\rm erf}
\Bigr(  \frac{y_{h' m'}}{\sqrt{2\sigma^2 x_{h' m'}}}  \Bigr)
\nn \\
&& \hspace{3cm} +
\sqrt{\frac{2 \sigma^2}{\pi x_{h' m'}}}\frac{y_{h' m'}}{x_{h' m'}} 
\exp\Bigr(-\frac{y_{h' m'}^2}{2\sigma^2x_{h' m'}}\Bigr)
\Biggr\} \frac{\partial y_{h'm'}}{\partial \gamma_{hm}}
\nn \\
&=&
\frac{\sigma^2}{x_{h' m}} {\rm erf}\Bigr(  \frac{y_{h' m}}{\sqrt{2\sigma^2 x_{h' m}}}  \Bigr)
\frac{(\hat{ \Sigma}_{\bm B m}^{-1} )_{h' h} }{(\hat{ \Sigma}_{\bm B m}^{-1} )_{h' h'}} 
\delta_{m,m'}.
\end{eqnarray}
From this, the mean is obtained from (\ref{bmean}),
\begin{eqnarray}
\bar{b}_{hm}
&=&
\frac{1}{Z_B F}\Bigr(
\int_{{\mathbb R}^{HM}} b_{hm}f({\bm B})d{\bm B}
-
\int_{{\mathbb R}^{HM}}
 b_{hm} \sum_{h' ,m'} \frac{ |b_{h' m'}| }{k}   f({\bm B})d{\bm B} \Bigr)
\nn\\
&=&
\frac{1}{Z_B F}
\Bigr(
\sigma^2\frac{\partial }{\partial \gamma_{hm}} 
F
-
\frac{1}{k} 
\sum_{h',m'}
\sigma^4\frac{\partial^2}{\partial \gamma_{hm} \partial \gamma_{h' m'} }
\int_{{\mathbb R}}
{\rm sign}(b_{h'm'}) \tilde{F}_{h'm'} 
db_{h m'}
\Bigr)
\nn\\
&=&
\frac{1}{Z_B F}
\Bigr[ 
\Bigr( \sum_{m'} (\hat{ \Sigma}_{\bm B m}^{-1} )_{hh} \gamma_{hm'} \Bigr) F
\nn \\
&& \hspace{2cm}
-
\frac{1}{k} \sum_{h',m'}
\Bigr\{
\frac{\sigma^2}{x_{h' m}}
{\rm erf}\Bigr(\frac{y_{h' m}}{\sqrt{2\sigma^2 x_{h' m}}} \Bigr) 
\frac{(\hat{ \Sigma}_{\bm B m}^{-1})_{h' h} }{(\hat{ \Sigma}_{\bm B m}^{-1})_{h' h'}} 
\delta_{m,m'} F
\nn \\
&& \hspace{5cm} + 
\epsilon_{h' m'}
\Bigr( \sum_{m''} (\hat{ \Sigma}_{\bm B m}^{-1})_{hh} \gamma_{hm''} \Bigr) F
\Bigr\}
\nn \\
&=&
\Bigr( \sum_{m'} (\hat{\Sigma}_{\bm B m}^{-1})_{hh} \gamma_{hm'} \Bigr)
-\sum_{h'}
\frac{\sigma^2 (\hat{ \Sigma}_{\bm B m}^{-1})_{h' h}}{k Z_B}
 {\rm erf}(\omega_{h' m}),
\end{eqnarray}
where $\omega_{hm} = y_{hm} / \sqrt{2 \sigma^2 x_{hm} }$. 
The equation (\ref{finalZB}) and the definition (\ref{x}) are used in the last line.
Substituting (\ref{defgamma}), one has the result in the main text.

\subsection{Evaluation of variance}
First one defines the quantity in (\ref{bvariance}),
\begin{eqnarray}
S_{hm}
&:=&
\frac{1}{Z_B F}\Bigr(
\int_{{\mathbb R}^{HM}} b_{hm}^2f({\bm B})d{\bm B}
- 
\frac{1}{k}
\int_{{\mathbb R}^{HM}} b_{hm}^2 \sum_{h' ,m'} |b_{h' m'}| f({\bm B})d{\bm B}\Bigr)
\nn\\
&=&
\frac{\sigma^4}{Z_B F}
\frac{\partial^2}{\partial \gamma_{hm}^2}
F
-
\frac{\sigma^6}{kZ_B F}\sum_{h' ,m'} 
\frac{\partial^3}{\partial \gamma_{hm}^2\partial \gamma_{h' m'}}
\int_{{\mathbb R}} {\rm sign}(b_{h' m'}) \tilde{F}_{h' m'} db_{h' m'}. 
\nn \\
\end{eqnarray}
The terms on r.h.s. are evaluated as
\begin{eqnarray}
\sigma^4
\frac{\partial^2}{\partial \gamma_{hm}^2} F
&=& \Bigr\{ \sigma^2 (\hat{\Sigma}_{\bm B m}^{-1} )_{hh}
+ \Bigr( \sum_{m''} (\hat{ \Sigma}_{\bm B m}^{-1} )_{hh} \gamma_{hm''} \Bigr)^2 \Bigr\} F,
\end{eqnarray}
\begin{eqnarray}
&& \sigma^6
\frac{\partial^2  }{\partial \gamma_{hm}^2}
\Bigr(
\frac{\partial }{\partial \gamma_{h' m'}}
\int_{{\mathbb R}} {\rm sign} (b_{h' m'}) \tilde{F}_{h' m'}db_{h' m'}
\Bigr) \nn \\
&=&  \sigma^2
\frac{\partial  }{\partial \gamma_{hm}}
\Bigr\{  \sigma^4 \frac{\partial  }{\partial \gamma_{hm}} 
\Bigr( \frac{\partial  }{\partial \gamma_{h'm'}}
\int_{{\mathbb R}} {\rm sign} (b_{h' m'}) \tilde{F}_{h' m'}db_{h' m'} \Bigr) \Bigr\}
\nn \\
&=&
\sigma^2\frac{\partial}{\partial \gamma_{hm}}
\Bigr\{
\frac{\sigma^2}{x_{h' m}}
{\rm erf}\Bigr(\frac{y_{h' m}}{\sqrt{2\sigma^2 x_{h' m}}} \Bigr) 
\frac{(\hat{ \Sigma}_{\bm B m}^{-1} )_{h' h} }{(\hat{\ \Sigma}_{\bm B m}^{-1} )_{h' h'}} 
\delta_{m,m'} F \nn \\
&& 
\hspace{5cm} +
\epsilon_{h' m'}
\Bigr( \sum_{m''} (\hat{\Sigma}_{\bm B m}^{-1} )_{hh} \gamma_{hm''} \Bigr) F \Bigr\}
\nn\\
&=& \Bigr\{ \frac{\sigma^2}{x_{h'm}} 
\sqrt{ \frac{2\sigma^2}{\pi x_{h'm}} }
\exp\Bigr(-\frac{y_{h' m}^2}{2\sigma^2 x_{h' m}} \Bigr)
\Bigr( \frac{(\hat{ \Sigma}_{\bm B m}^{-1} )_{h' h} }{(\hat{ \Sigma}_{\bm B m}^{-1} )_{h' h'}} \Bigr)^2
\delta_{m,m'} \nn \\
&& + 
2 \frac{\sigma^2}{x_{h' m}}
{\rm erf}\Bigr(\frac{y_{h' m}}{\sqrt{2\sigma^2 x_{h' m}}} \Bigr) 
\frac{(\hat{ \Sigma}_{\bm B m}^{-1} )_{h' h} }{(\hat{ \Sigma}_{\bm B m}^{-1} )_{h' h'} } 
\delta_{m,m'} 
\Bigr( \sum_{m''} (\hat{ \Sigma}_{\bm B m}^{-1} )_{hh} \gamma_{hm''} \Bigr)
\nn\\
&& + 
\sigma^2 \epsilon_{h'm'}
(\hat{ \Sigma}_{\bm B m}^{-1} )_{hh}
+ \epsilon_{h'm'}
\Bigr( \sum_{m''} (\hat{ \Sigma}_{\bm B m}^{-1} )_{hh}\gamma_{hm''} \Bigr)^2
\Bigr\} F.
\end{eqnarray}
Hence
\begin{eqnarray}
S_{hm}
&=&
\frac{1}{Z_B F} \Bigr( 1 - \frac{1}{k} \sum_{h'm'} \epsilon_{h'm'} \Bigr) 
\Bigr\{ \sigma^2 (\hat{ \Sigma}_{\bm B m}^{-1} )_{hh}
+ \Bigr( \sum_{m''} (\hat{ \Sigma}_{\bm B m}^{-1} )_{hh} \gamma_{hm''} \Bigr)^2 \Bigr\}F
\nn \\
&& - \frac{1}{kZ_B F} \sum_{h'}
\Bigr\{
\frac{\sigma^2}{x_{h'm}} 
\sqrt{ \frac{2\sigma^2}{\pi x_{h'm}} }
\exp\Bigr(-\frac{y_{h' m}^2}{2\sigma^2 x_{h' m}} \Bigr)
\Bigr( \frac{(\hat{ \Sigma}_{\bm B m}^{-1} )_{h' h} }{(\hat{ \Sigma}_{\bm B m}^{-1} )_{h' h'} } \Bigr)^2 \nn \\
&& \hspace{1.2cm} + 
2 \frac{\sigma^2}{x_{h' m}}
{\rm erf}\Bigr(\frac{y_{h' m}}{\sqrt{2\sigma^2 x_{h' m}}} \Bigr) 
\frac{(\hat{\Sigma}_{\bm B m}^{-1} )_{h' h} }{(\hat{\Sigma}_{\bm B m}^{-1} )_{h' h'} } 
\Bigr( \sum_{m''} (\hat{\Sigma}_{\bm B m}^{-1} )_{hh} \gamma_{hm''} \Bigr)
\Bigr\}F
\nn \\
&=&
\Bigr\{ \sigma^2 (\hat{ \Sigma}_{\bm B m}^{-1} )_{hh}
+ \Bigr( \sum_{m''} (\hat{ \Sigma}_{\bm B m}^{-1} )_{hh}\gamma_{hm''} \Bigr)^2 \Bigr\}
\nn \\
&& - \frac{\sigma^2}{k Z_B}\sum_{h'}
\left\{
\sqrt{ \frac{2\sigma^2}{\pi (\hat{ \Sigma}_{\bm B m}^{-1} )_{h' h'}} }
\exp( -\omega_{h' m}^2 )
\Bigr( (\hat{ \Sigma}_{\bm B m}^{-1} )_{h' h} \Bigr)^2 \right. \nn \\
&& \left. \hspace{2.3cm} + 
 2 (\hat{\Sigma}_{\bm B m}^{-1} )_{h' h} {\rm erf} (\omega_{h' m} ) 
\Bigr( \sum_{m''} (\hat{\Sigma}_{\bm B m}^{-1} )_{hh} \gamma_{hm''} \Bigr)
\right\},
\end{eqnarray}
where the definitions (\ref{x}) and (\ref{y}) are used.
After insertion of this result into covariance matrix, one has from (\ref{bvariance}),
\begin{eqnarray}
( \Sigma_{{\bm B}m})_{hh}
&=&
S_{hm} - \bar{b}_{hm}^2
\nn\\
&=&
\sigma^2 (\hat{ \Sigma}_{\bm B m}^{-1} )_{hh}
\nn \\
&&
-
\sum_{h'}
\sqrt{\frac{2}{\pi \sigma^2 (\hat{ \Sigma}_{\bm B m}^{-1} )_{h'h'} }  }
\frac{\{ \sigma^2 (\hat{ \Sigma}_{\bm B m}^{-1} )_{h' h} \}^2}{k Z_B}
\exp(-\omega_{h' m}^2 )
\nn\\
&& -
\Bigr(
\frac{\sigma^2}{k Z_B}
\sum_{h'}
(\hat{ \Sigma}_{\bm B m}^{-1} )_{h' h}{\rm erf}(\omega_{h' m}) 
\Bigr)^2.
\end{eqnarray}




\section*{References}

\begin{thebibliography}{99}

\bibitem{SRJ}
Srebro N, Rennie J~D~M and Jaakkola T~S 2005 {\em Advances in Neural
  Information Processing System\/} {\bf 17} 1329--1336

\bibitem{CP}
Cand\`{e}s E~J and Plan Y 2010 {\em Proceedings of the IEEE\/} {\bf 98}
  925--936

\bibitem{CCS}
Cai J~F, Cand\`{e}s E~J and Shen Z 2010 {\em SIAM Journal on Optimization\/}
  {\bf 20} 1956--1982

\bibitem{SM}
Salakhutdinov R and Mnih A 2008 {\em Proceedings of the 25th International
  Conference on Machine Learning\/}  880--887

\bibitem{KKMSZ}
Kabashima Y, Krzakala F, M\'{e}zard M, Sakata A and Zdeborov\'{a} L 2016 {\em
  IEEE Transactions on Information Theory\/} {\bf 62} 4228 --4265

\bibitem{NS}
Nakajima S and Sugiyama M 2011 {\em Journal of Machine Learning Research\/}
  {\bf 12} 2583--2648

\bibitem{NSBT}
Nakajima S, Sugiyama M, Babacan S~D and Tomioka R 2013 {\em Journal of Machine
  Learning Research\/} {\bf 14} 1--37

\bibitem{OF1}
Olshausen B~A and Field D~J 1996 {\em Nature\/} {\bf 381} 607--609

\bibitem{OF2}
Olshausen B~A and Field D~J 1997 {\em Vision Research\/} {\bf 37} 3311--3325

\bibitem{EAH}
Engan K, Aase S~O and Husoy J~H 1999 {\em Proceedings of IEEE International
  Conference of Acoustics, Speech, and Signal Processing\/}  2443--2446

\bibitem{AEB}
Aharon M, Elad M and Bruckstein A~M 2006 {\em IEEE Transactions on Signal
  Processing\/} {\bf 54} 4311--4322

\bibitem{ZT}
Zhou T and Tao D 2011 {\em Proceedings of the 28th International Conference on
  Machine Learning\/}  33--40

\bibitem{SK}
Sakata A and Kabashima Y 2013 {\em EPL\/} {\bf 103} 28008

\bibitem{KMZ}
Krzakala F, M\'{e}zard M and Zdeborov\'{a} L 2013 {\em Proceedings of IEEE
  International Symposium on Information Theory\/}  659 -- 663

\bibitem{ref9}
Bishop C~M 2006 {\em Pattern Recognition and Machine Learning\/} (Springer)

\bibitem{NW}
Nakajima S and Watanabe S 2007 {\em Neural Computation\/} {\bf 19} 1112--1153

\end{thebibliography}
\end{document}